\documentclass[journal]{vgtc}          

\usepackage{mathptmx}
\usepackage{graphicx}
\usepackage{times}
\usepackage{amssymb}
\usepackage{subfigure}
\usepackage{url}
\usepackage{amsmath}
\usepackage{bm}
\usepackage{color}
\usepackage{multicol}
\usepackage{multirow}
\usepackage{dcolumn}

\newcommand{\ignore}[1]{}

\newcommand{\ie}{\emph{i.e., }}
\newcommand{\etal}{\emph{et al.}}
\newcommand{\Poincare}{Poincar\'{e} }
\newcommand{\Avizo}{Avizo${}^{\text{\textregistered}}$\hspace{1pt}}

\newcommand{\dv}{$\Delta V$ }
\newcommand{\dvnospace}{$\Delta V$}

\newcommand{\crtbp}{CR3BP }
\newcommand{\crtbpNoSpace}{CR3BP}
\newcommand{\pcrtbp}{PCR3BP }
\newcommand{\pcrtbpNoSpace}{PCR3BP}
\newcommand{\hide}[1]{}
\newcommand{\eg}{\emph{e.g.}, }

\newcommand{\fig}[1]{Fig.~\ref{#1}}

\providecommand{\e}[1]{\ensuremath{\times 10^{#1}}}

\newcommand{\dro}{$p=3$ DRO }
\newcommand{\droNS}{$p=3$ DRO}

\usepackage[bookmarks, backref=true, linkcolor=black,  colorlinks=true,
  citecolor= black,
  pageanchor=true,
  urlcolor = black,
  plainpages = false,
  linktocpage]{hyperref} 
\hypersetup{
  pdfauthor = {Schlei, Tricoche, Howell},
  pdftitle = {Extraction and Visualization of \Poincare Map Topology for Spacecraft Trajectory Design},
  pdfsubject = {IEEE VIS},
  pdfkeywords = {Trajectory planning and design, \Poincare map, dynamical systems, topology extraction, invariant manifolds, chaos}
  colorlinks=true,
  linkcolor= black,
  citecolor= black,
  pageanchor=true,
  urlcolor = black,
  plainpages = false,
  linktocpage
}

\onlineid{0}

\vgtccategory{Application}

\vgtcinsertpkg

\graphicspath{{./figs/}}

\title{Extraction and Visualization of \Poincare Map Topology for Spacecraft Trajectory Design}

\author{Xavier Tricoche, Wayne Schlei, and Kathleen C. Howell}
\authorfooter{
\item
Xavier Tricoche is an Associate Professor of Computer Science at Purdue University, e-mail: xmt@purdue.edu.
\item
Wayne Schlei is a Mission Design Engineer, JHU Applied Physics Lab, e-mail: Wayne.Schlei@jhuapl.edu.
\item
Kathleen Howell is the Hsu-Lo Professor of Aeronautics and Astronautics at Purdue University, email: howell@purdue.edu.
}

\shortauthortitle{Tricoche \MakeLowercase{\textit{et al.}}}

\abstract{Mission designers must study many dynamical models to plan a low-cost spacecraft trajectory that satisfies mission constraints. They routinely use \Poincare maps to search for a suitable path through the interconnected web of periodic orbits and invariant manifolds found in multi-body gravitational systems. This paper is concerned with the extraction and interactive visual exploration of this structural landscape to assist spacecraft trajectory planning. We propose algorithmic solutions that address the specific challenges posed by the characterization of the topology in astrodynamics problems and allow for an effective visual analysis of the resulting information. This visualization framework is applied to the circular restricted three-body problem (CR3BP), where it reveals novel periodic orbits with their relevant invariant manifolds in a suitable format for interactive transfer selection. Representative design problems illustrate how spacecraft path planners can leverage our topology visualization to fully exploit the natural dynamics pathways for energy-efficient trajectory designs.}


\CCScatlist{ 
 \CCScat{J.2.1}{Physical Sciences and Engineering}%
{Aerospace Engineering}{};
}

\teaser{
  \centering

    \includegraphics[width=\columnwidth]{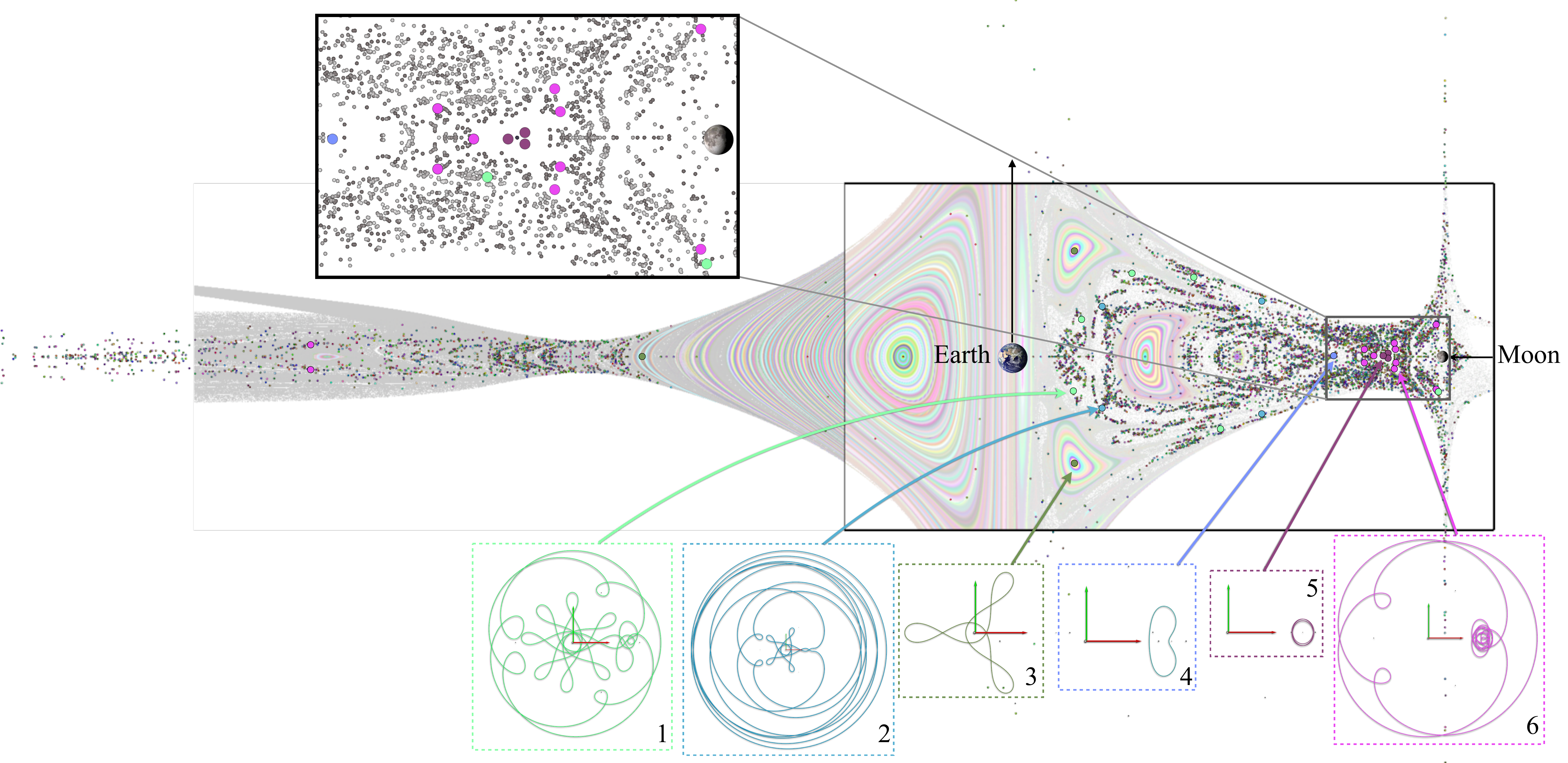}
    \caption{An overview of the 1450 periodic orbits (most of which previously unknown) extracted by the proposed method in the \Poincare map of the Earth-Moon three-body system at an energy level that enables spacecraft motion across the entire phase space. A sample of orbits shows the corresponding trajectories around the Earth and the Moon. Our method allows spacecraft trajectory designers to visually find and evaluate natural connections between these orbits as they seek energy-efficient paths that satisfy mission constraints.
    }
    \label{sf:emfpFull}
  \label{fig:Teaser}

}

\begin{document}

\firstsection{Introduction}

\maketitle

The design of a spacecraft trajectory is key to the success of any space mission. The spacecraft path must deliver the scientific objectives of the mission under the constraints imposed by the laws of physics and a targeted mission price. Although many factors influence the mission cost, the price is strongly driven by the spacecraft mass. Beside the spacecraft operational equipment (\eg{} antenna, engines, solar arrays, etc...), this mass is made up of \emph{payload} and \emph{propellant}. The payload is the collection of scientific instruments required to acquire the mission data. Course corrections or maneuvers are accomplished by performing a change in spacecraft velocity (called \dvnospace) while expelling propellant. Although propellant is necessary to perform maneuvers, payload mass is typically favored over propellant mass. Such a trade-off produces more scientific return for the mission and may reduce its cost. The role of a spacecraft trajectory designer is therefore to devise a pathway that minimizes the amount of propellant required to transport the vehicle to mission objectives.

An essential aspect of the path planner's work consists in finding suitable trajectory candidates within relevant multi-body gravitational models. Specifically, designers strive to exploit the dynamics that naturally exists in the model to control the spacecraft's path while requiring little or no maneuvers. Widely used in practice, the \emph{circular restricted three-body problem} (CR3BP) offers a versatile model in which the gravitational field is defined by two massive bodies (the Sun and a planet or a planet and a moon). The CR3BP exhibits a rich dynamics, and lends itself to a realistic simulation of the spacecraft behavior during a planned mission. Since the corresponding design space, the \emph{phase space}, is high-dimensional, the dynamics is preferrably studied on a properly chosen planar surface utilizing the \emph{\Poincare map} (or first-return mapping). The \Poincare map offers a phase-space snapshot of all crossing trajectories while reducing the dimension of the problem. \fig{sf:emfpFull} shows a \Poincare map of the Earth-Moon CR3BP in which periodic trajectories have been highlighted. Yet, even in this more tractable setup, the analysis of the CR3BP remains a difficult task.
Knowing the topological skeleton of the \Poincare map, which indicates how orbital structures are naturally interconnected, would supply spacecraft path-planning with a plethora of options and pathways that save on propellant usage. Unfortunately, the automatic extraction of the CR3BP's \Poincare map topology is challenging due to numerical sensitivities, the extensive presence of chaotic dynamics, and violated assumptions in the construction of the map. As a result, the topological insight that has been used so far to design low-cost transfers remains very limited. Instead, spacecraft trajectory designers usually probe \Poincare maps in basic puncture plots, which they then use to initiate the search for relevant connections through numerical optimization, a process that can be ineffective and tedious.

In this context, this paper makes the following contributions. First, we present a number of significant technical solutions to the extraction of the \Poincare map topology that decisively improve upon existing methods in the planar CR3BP. Second, we propose an interactive visualization that delivers this topological information in an actionable visual form to the space trajectory designer. Specifically, we record the hierarchical make-up of the separatrices of the topology, which allows designers to instantaneously navigate these manifolds up and down stream at no additional computational cost as they explore the visualization. We demonstrate the effectiveness of our solution in the context of representative design scenarios, in which our topology visualization enables the rapid definition of free-flowing connections between distant spatial locations.

\section{Related Work}\label{sec:prior work}

The three-body problem defines a vector field and the topological structure of the \crtbp that we aim to visualize in a \Poincare map is the discrete signature of a three-dimensional vector field topology. There exists an abundant literature in the visualization community focused on the use of topological methods to represent vector fields~\cite{Laramee:2005:Topology-Based,Pobitzer:2011:The-state,Heine:2016:A-survey}. In the discrete realm, L\"offelmann \etal{} introduced multiple methods for the  visualization of maps~\cite{Loffelmann:1997:Stream,Loffelmann:1998:Enhancing,Loffelmann:1998:Visualizing,
Loffelmann:1998:VisualizingP}, thereby emphasizing the continuous structures that control the dynamics. This work followed the seminal contributions of Abraham and Shaw to the expressive visualization of dynamical systems~\cite{Abraham:1992:Dynamics}. Peikert and Sadlo applied a \Poincare map approach to the visualization of vortex rings in a flow recirculation bubble~\cite{Peikert:2007:Visualization,Peikert:2008:Flow}. Their work introduced a method to find the location of center-type fixed points and revealed the  tangles formed by the separatrices of the topology. Tricoche  \etal{} proposed to compute the finite-time Lyapunov exponent (FTLE)~\cite{Haller:2001:Distinguished} of the \Poincare map to obtain an image-based visualization of separatrices in area-preserving maps~\cite{Tricoche:2012:Visualizing}.
Sanderson \etal{} presented a method to extract the geometry of the island chains present in \Poincare maps~\cite{Sanderson:2010:Analysis} from the successive returns of randomly selected streamlines. They also described a method to infer the location of fixed points from that geometry. Tricoche \etal{} later proposed a method to extract the fixed points of area-preserving maps along with their associated separatrices~\cite{Tricoche:2011:Visualization} and applied it to magnetic confinement in a fusion reactor. Sagrista \etal{} introduced an extension of FTLE to inertial dynamics and applied it to the analysis of this dynamic in high-dimensional phase space~\cite{Sagrista:2016:Topological}.

The visualization of the \Poincare map has also been considered in the context of astrodynamics and space mission design applications. Short \etal{} applied the concept of Lagrangian coherent structures~\cite{Haller:2001:Distinguished} to the visualization of the \Poincare map in the \crtbp and other multi-body systems~\cite{Short:2011:Lagrangian,Short:2014:Lagrangian}.
Haapala \etal{} presented a glyph-based solution for the visualization of four-dimensional \Poincare maps in the study of the spatial three-body problem~\cite{Haapala:2014:Representations}. Davis \etal{} proposed a visualization method for multidimensional \Poincare maps to identify design options around Saturn's poles~\cite{Davis:2018:Trajectory}.
Most closely related to the present work, Schlei \etal{}~\cite{Schlei:2014:Enhanced} presented several modifications to the fixed point extraction method of Tricoche \etal{}~\cite{Tricoche:2011:Visualization} to improve its results in multi-body gravitational environments. We discuss their work in more detail in the presentation of our algorithm in \autoref{sec:algorithm}.

Unfortunately, none of the techniques presented to date are able to capture the complex topology of the \crtbpNoSpace, nor do they support the kind of visual analysis of the topology that spacecraft trajectory designers need.

\section{Application Background}

We provide in this section a brief introduction to space trajectory design, which provides the application context and the motivation for the work presented in the remainder of this paper. We start with a short presentation of the circular restricted three-body problem in its planar form before discussing its analysis through \Poincare maps.

\subsection{Planar Circular Restricted Three-Body Problem}\label{sec:cr3bp}

\begin{figure}[hbt]
    \includegraphics[width=0.99\columnwidth]{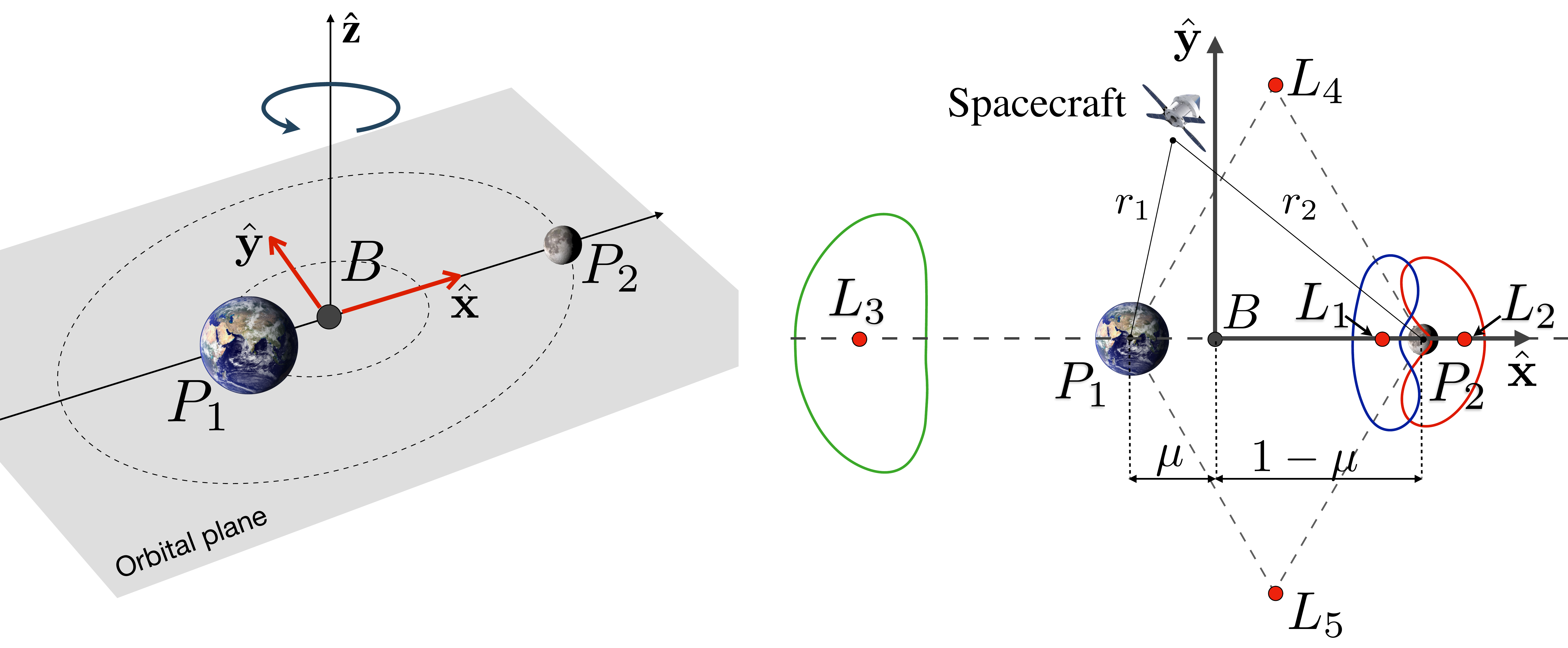}
    \caption{The planar circular restricted three-body problem. Left: Rotating coordinate frame and orbital plane. Right: Spacecraft with primaries, Lagrange points (in red), and Lyapunov orbits surrounding $L_1$ (blue), $L_2$ (red), and $L_3$ (green), in the orbital plane.}
    \label{fig:cr3bp_explanation}
\end{figure}

The gravitational model considered in this paper describes the motion of a spacecraft under the influence of two celestial bodies that form an orbital system (\eg Earth and Moon or Sun and Earth).  The motion of a spacecraft under the influence of the combined gravitational field is then modeled as the circular restricted three-body problem (\crtbpNoSpace).  Assume a pair of gravitating bodies, $P_1$ and $P_2$ with corresponding masses $m_1>m_2$ and associated \textit{gravity parameter} $\mu = m_2/(m_1+m_2)$, revolve around their common barycenter $B$ in circular orbits (see \autoref{fig:cr3bp_explanation}, left). We further assume that the spacecraft resides in the \emph{orbital plane} (that is $z\equiv 0$), in which case the \crtbp equations are reduced to the planar \crtbp (PCR3BP).
Position and speed of the spacecraft are then given by its \textit{state vector} $\mathbf{x}=[x,y,\dot{x},\dot{y}]^T$ (with $\dot{a}$ denoting the time derivative $\tfrac{da}{dt}$). Coordinates are expressed in a rotating reference frame with the origin at the barycenter, the axis $\hat{\mathbf{x}}$ aligned with the $\overrightarrow{P_1P_2}$ line and $\hat{\mathbf{y}}$ is aligned with the velocity vector of $P_2$ with respect to $P_1$.

Using so-called \emph{nondimensional units} of distance, mass, and time\footnote{For reference, a nondimensional position unit in the Earth-Moon (\textit{EM}) system is equivalent to 384388.174 km whereas a nondimensional velocity unit is 1.02456261 km/s.~\cite{Danby:1992:Fundamentals}}, a pseudo-potential function $\Upsilon$ can be defined as
\begin{equation}
\Upsilon(x,y) = \frac{1-\mu}{r_1} + \frac{\mu}{r_2} + \frac{1}{2}(x^2+y^2),
\end{equation}
whereby, $r_1$ (resp. $r_2$) denotes the distance from the spacecraft to $P_1$ (resp. $P_2$). Note that $\Upsilon$ adds to the standard gravitational potential a third term that accounts for the rotation of the reference frame. The \pcrtbp model is then a dynamical system described by its \emph{equations of motion}:
\begin{equation}\label{eq:CR3BP}
\ddot{x} - 2\dot{y} = \frac{\partial\Upsilon}{\partial x} \text{ and }
\ddot{y} + 2\dot{x} = \frac{\partial\Upsilon}{\partial y},
\end{equation}
which defines an ordinary differential equation (ODE) on the state vector $\mathbf{x}$. Solutions to this ODE are integral curves that are referred to as \emph{orbits}. The energy of the system is characterized by its \textit{Jacobi constant} $C$, which is defined as:
\begin{equation}\label{eq:jc}
C = 2\Upsilon(x,y) - (\dot{x}^2+\dot{y}^2),
\end{equation}
with the total planar velocity $V = (\dot{x}^2+\dot{y}^2)^{\frac{1}{2}}$~\cite{Danby:1992:Fundamentals}.
This system is conservative (the potential and kinetic energy remains constant) and the associated flow, which controls the trajectory of the spacecraft, is area-preserving~\cite{Lichtenberg:1992:Regular}. $C$ is invariant under the action of the \pcrtbpNoSpace. In astrodynamics, however, $C$ can be modified through spacecraft maneuvers (\dvnospace), which use propellant to change $V$. Observe that larger $C$ values correspond to \emph{lower} energy levels and vice versa.

Some particular solutions to the equations of motion play an important role in astrodynamics. The five \emph{Lagrange points} are equilibrium solutions of \autoref{eq:CR3BP} in the rotating frame, that is $\nabla\Upsilon = \mathbf{0}$. \emph{Lyapunov orbits} are planar periodic orbits about the Lagrange points. Refer to \autoref{fig:cr3bp_explanation}, right.
More generally, periodic orbits (which are \emph{closed} streamlines) play a central role in the design process~\cite{Gomez:2004:Connecting,Howell:2001:Families}. For instance, a periodic orbit allows a satellite to hover in an ideal vicinity for observations and data collection without constant corrections. Periodic orbits can also be used to design trajectories that offer ideal observation solutions with respect to the sun light, facilitate transfers between different trajectories, or provide long term parking options for spacecraft disposal at the end of a mission. An important periodic orbit in practice is the distant retrograde orbit (DRO), which performs counterclockwise rotations around the smaller body (\eg{}, the Moon).

Finally, the following terms are used repeatedly in the design scenarios that we present in \autoref{sec:application}. The term \emph{capture} refers to the access of an orbit around a celestial body. A \emph{ballistic capture} occurs when capture is enabled by gravity and does not require a maneuver.

\subsection{Topological Structure in \Poincare Maps} \label{sec:background}\label{sec:theory}\label{sec:map}

This section reviews basic theoretical results about area-preserving maps and \Poincare maps that apply to the PCR3BP. Additional details can be found in classical references~\cite{Guckenheimer:1983:Nonlinear,Lichtenberg:1992:Regular}.

A dynamical system associated with a vector field $\mathbf{v}$ defines a \emph{flow map} $\mathbf{x}_f$ with $\dot{\mathbf{x}}_f = \mathbf{v}(\mathbf{x}_f)$ such that $\mathbf{x}_f(t, t_0, \mathbf{x}_0)$ describes the transport from an initial \textit{state} $\mathbf{x}_0$ at time $t_0$ to its later state at time $t$.

\begin{figure}[hbt]
   \centering
   $\vcenter{\hbox{\subfigure[CR3BP and \Poincare map]{\label{sf:PMapDef}\label{fig:poincare_map}\includegraphics[width=0.49\columnwidth]{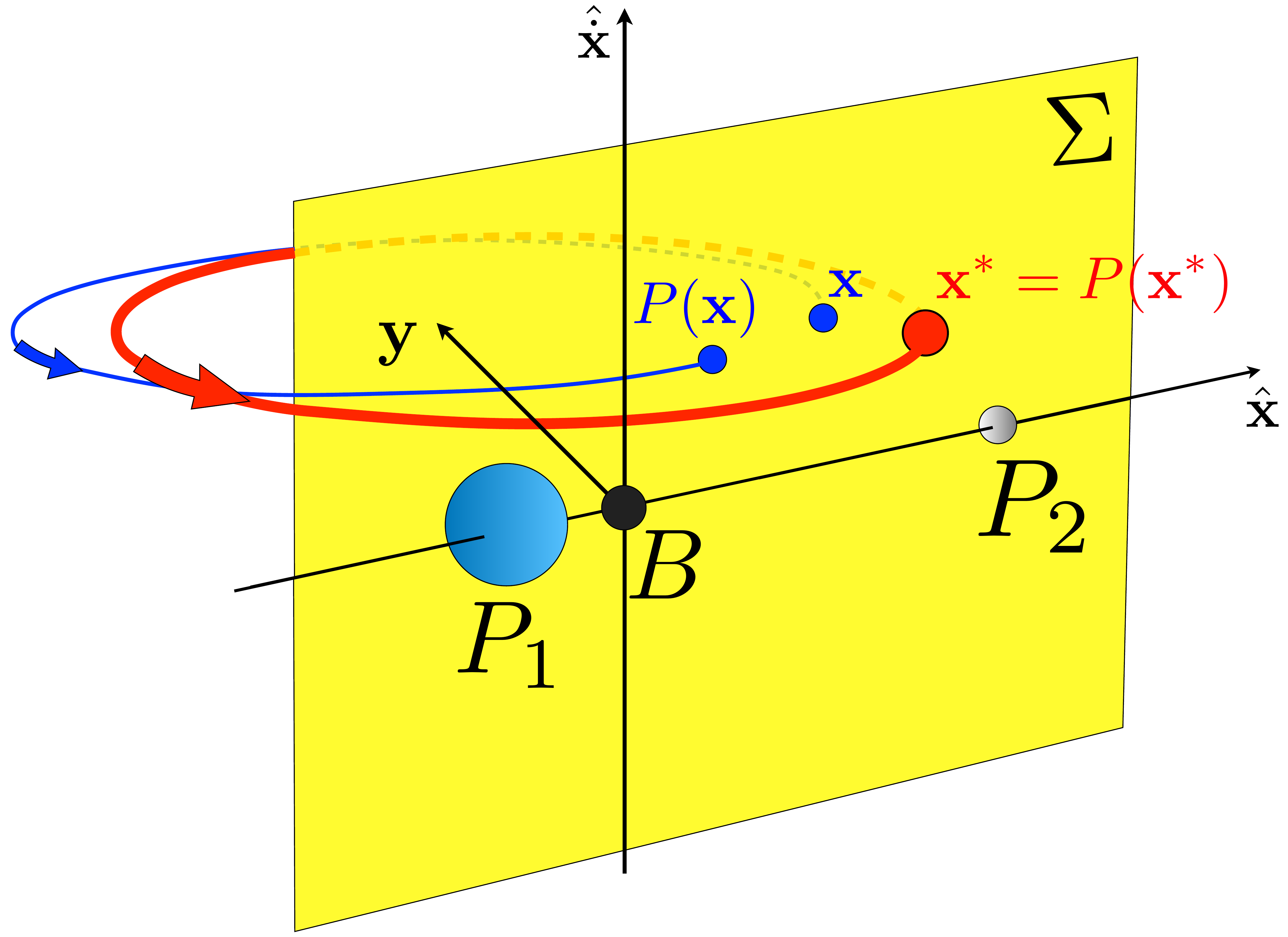}}}}$
   $\vcenter{\hbox{\subfigure[\Poincare map topology]{\label{fig:topology}\includegraphics[width=0.49\columnwidth]{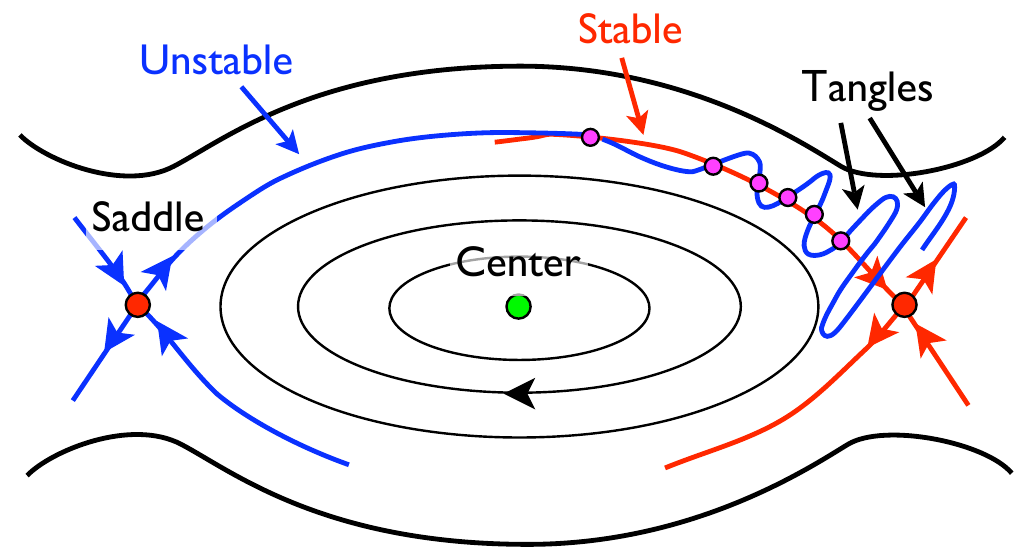}}}}$
      \caption{\Poincare map in PCR3BP and associated topological structure.
	 }
   \label{fig:topoStructure}
\end{figure}

Let $\varSigma$ represent a plane that is transverse\footnote{The plane is transverse if the vector field is nowhere tangent to $\varSigma$} to the flow and let $\mathbf{x}_0$ be an initial state on $\varSigma$. The \Poincare map, or the first-return map $\bm{\mathit{P}}(\mathbf{x}_0) := \mathbf{x}_0 \mapsto \bm{\mathit{P}}_\Sigma(\mathbf{x}_0)$ corresponds to the first crossing of $\varSigma$ by the trajectory starting at $\mathbf{x}_0$.
Multiple iterations of the \Poincare map are then computed by compounding the first return map, \eg{}
$\bm{\mathit{P}}^p(\mathbf{x}_0) = \bm{\mathit{P}}_\Sigma(\bm{\mathit{P}}_\Sigma( \ldots \bm{\mathit{P}}_\Sigma(\mathbf{x}_0)))$
for $p$ returns.  Both the initial state and first return to $\varSigma$ are shown on the yellow plane in \autoref{sf:PMapDef}. Later sections refer to the vector-valued (displacement) mapping $\bm{\Delta}_p$ associated with a given period $p$, which is defined as:
\begin{equation}\label{eq:delta}
\bm{\Delta}_p(\mathbf{x})=\bm{\mathit{P}}^p(\mathbf{x})-\mathbf{x}.
\end{equation}

Three dynamic behaviors co-exist on a \Poincare map
of the planar \pcrtbpNoSpace: \emph{periodicity}, \emph{quasi-periodicity}, and \emph{chaos}. Visible in \autoref{sf:PMapDef}, a periodic state, $\mathbf{x}^*$, returns to the same state through the \Poincare map, \ie
$\bm{\Delta}_{p}(\mathbf{x}^{*}) = \bm{0}$,
where $p$ represents the number of iterations required for a $p$-periodic trajectory to complete an orbit. These $p$ distinct returns are called \textit{fixed points} of the \Poincare map and, owing to the \emph{aera-preserving} nature of the \pcrtbpNoSpace, they can either be \emph{centers} or \emph{saddles}.
Refer to \autoref{fig:topology}.
So-called \emph{stable} and \emph{unstable manifolds} (known as \emph{separatrices} in vector field topology) emerge from the saddle fixed points and flow into and out of the periodic orbits, respectively. A fundamental feature of \Poincare map topology is the existence of connections between saddle fixed points, in which stable and unstable manifolds intersect an infinite number of times, creating  tangles as seen in \autoref{fig:topology}. Connections between saddle fixed points of the same periodic orbit are called \emph{homoclinic connections}, whereas those that join saddle fixed points of two distinct orbits are called \emph{heteroclinic connections}.

It can be shown that, for a given value of the Jacobi constant $C$, the dynamics of the \pcrtbp is confined to a torus~\cite{Strogatz:1994:Nonlinear},
in which the motion is then characterized by the so-called \emph{winding number}
 $w = \frac{\omega_1}{\omega_2}$,
where $\omega_1$ and $\omega_2$ are the \emph{poloidal} (measured in small circle) and \emph{toroidal} (measured in large circle) rotation frequencies, respectively.  The winding number permits us to classify trajectories: rational numbers $w=\frac{q}{p}, p,q \in \mathbb{N^*},p\mbox{ and }q\mbox{ mutually prime}$ correspond to periodic orbits. In this case, $q$ corresponds to the number of poloidal rotations performed during $p$ toroidal rotations and $p$ is the period of the fixed point. In contrast, quasi-periodic trajectories possess irrational winding numbers: a quasi-periodic orbit will never trace exactly the same path along the torus, and in the case of chaotic trajectories, the winding number is undefined.

In the vicinity of a fixed point of period $p$, the spatial derivative of the flow map $\mathbf{x}_f$ after $p$ returns is known as the \emph{monodromy} matrix $\mathcal{M}$.
The eigenvalues of $\mathcal{M}$ determine the type of a fixed point. Since the flow map is area-preserving, these eigenvalues come in reciprocal pairs. The distinction between stable, center, and unstable types is done as follows.
\begin{equation}\label{eq:stability}
    \left| \lambda_i \right| < 1 \mbox{: stable}, \\
    \left| \lambda_i \right| = 1 \mbox{: center}, \\
    \left| \lambda_i \right| > 1 \mbox{: unstable}.
\end{equation}
In the case of a saddle-type fixed point, one eigenvalue has a stable type while the other eigenvalue has an unstable type. The corresponding eigenvectors  ($\mathbf{e}_i$ with $i=S,U$ for stable and unstable, respectively) indicate the tangent of its invariant manifolds $W^i$.

An alternative stability classification is possible through a \emph{stability index},  $\nu_{SI}$, defined as
\begin{equation}\label{eq:SI}
\nu_{SI} = \tfrac{1}{2}(\text{tr}(\mathcal{M}) - 2).
\end{equation}
Orbits are unstable when $\lvert\nu_{SI}\rvert>1.0$ and stable otherwise.

\section{\Poincare Map Topology Extraction in the \pcrtbpNoSpace}\label{sec:contributions}\label{sec:algorithm}

As noted previously, no existing method can successfully extract the topology of the \Poincare map in the PCR3BP. Nonetheless, our work is closely related to the work of Schlei \etal{}~\cite{Schlei:2014:Enhanced}. In this section, we first summarize the main points of this existing approach and discuss its shortcomings. We then present our algorithmic contributions. Note that in the following we refer to a maximum period $p_{max}$ to consider. A maximum period is needed because the \pcrtbp possesses a \emph{fractal} topology that contains an \emph{infinite} number of periodic orbits and associated invariant manifolds for arbitrarily high values of the period. A maximum period is therefore a means to bound the scope of the topology extraction. From a practical standpoint, a maximum period also amounts to an upper bound on the time and distance that a spacecraft would need to travel to return to the same point.

\subsection{Existing Approach and Limitations}
\label{sec:existing approach}

Schlei \etal{} proposed a method to extract fixed points and separatrices of the \Poincare map topology~\cite{Schlei:2014:Enhanced}, which extends to the \pcrtbp a method proposed by Tricoche \etal{}~\cite{Tricoche:2011:Visualization}.
Its main input parameters are $p_{max}$: maximum period of the \Poincare map to consider and $G$: regular grid that covers the considered domain of the \Poincare section in the $(x,\dot{x}),\, y=0$ plane. It proceeds as follows.

\begin{figure}[hbt]
   \centering
   \includegraphics[width=0.99\columnwidth]{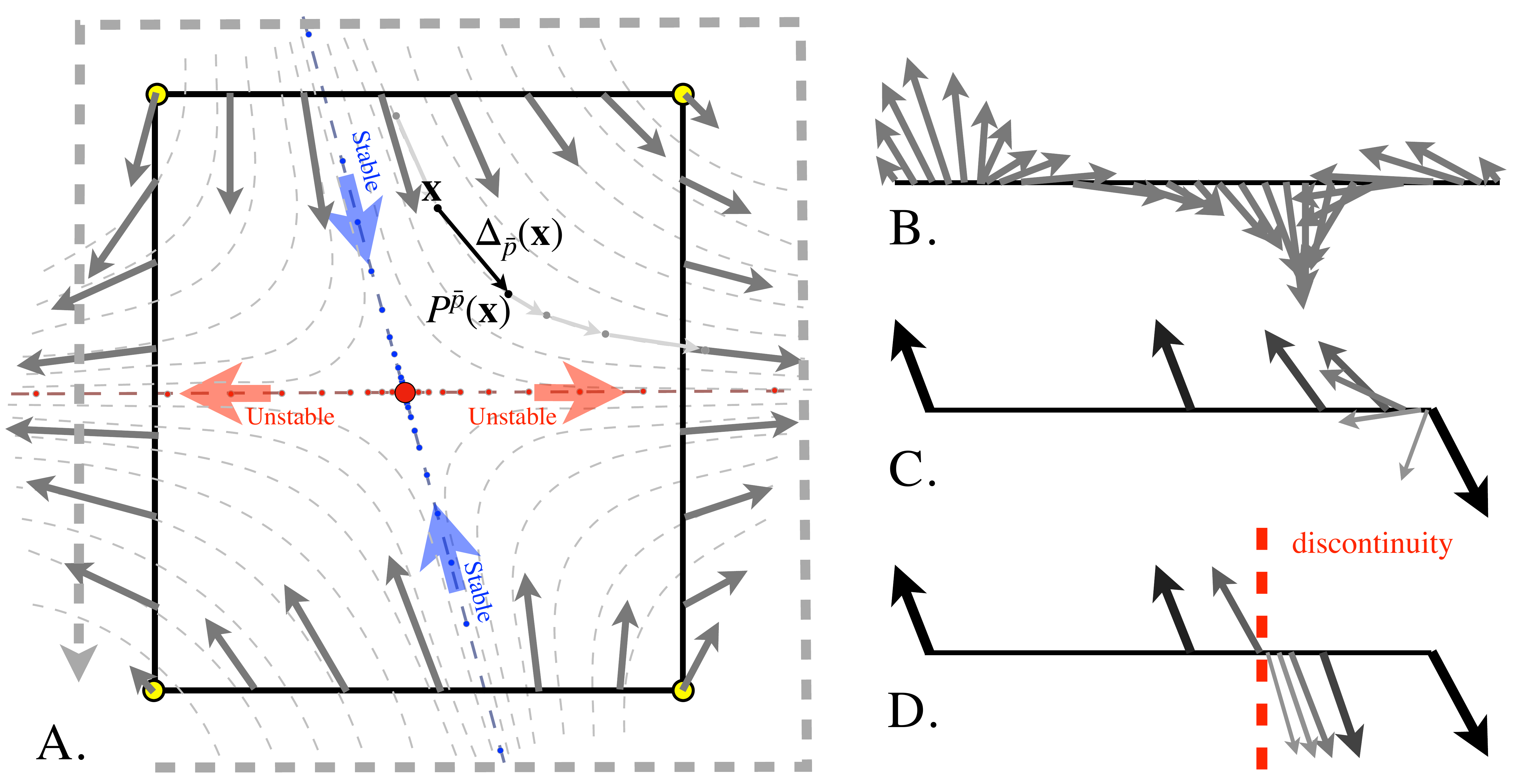}
   \caption{\Poincare index computation for $\bm{\Delta}_{\bar{p}}$. A. index computed along the edges of a cell containing a saddle fixed point of period $\bar{p}$. B. the corresponding $\bm{\Delta}_{\bar{p}}$ values complete 1 clockwise rotation (index -1). C. Nonlinearity in the rotation of $\bm{\Delta}_{\bar{p}}$ is resolved via adaptive sampling. D. Adaptive sampling fails to resolve orientation change when a transversality violation creates a discontinuity of $\bm{\Delta}_{\bar{p}}$.}
   \label{fig:poincare_index}
\end{figure}

\vspace{0.1cm}\noindent
\textbf{Map sampling:} Evaluate the \Poincare map, through numerical integration of the equation of motion (\autoref{eq:CR3BP}), at each vertex of the grid $G$, for a number of iterations $k \geq p_{max}$. For each computed orbit, record a vector $\mathbf{W}=(w_{xy}, w_{x\dot{x}}, w_{\dot{x},y})$ of 3 winding numbers (see \autoref{sec:theory}) in the $(x,\dot{x})$, $(x,y)$, and $(\dot{x},y)$ planes, respectively.
For each winding number $w$, compute a best rational approximation $w \approx q^*/p^*$ such that $p^* \leq p_{max}$ and store $p^*$. In general, the computed orbit will be quasi-periodic (see \autoref{sec:theory}) and $w$ will be an irrational number.

\vspace{0.1cm}\noindent
\textbf{Index computation:} For each cell of the grid $G$, consider the set of periods $p^*$ recorded at its 4 vertices. For each such period $\bar{p}$, compute via adaptive sampling the \Poincare index of the displacement map $\bm{\Delta}_{\bar{p}}$ (see \autoref{eq:delta}), \ie{} the number of counterclockwise rotations of $\bm{\Delta}_{\bar{p}}$, along the cell edges. Refer to \autoref{fig:poincare_index} for an illustration. Assuming a fine enough sampling\footnote{The cell must contain no more than one fixed point of the same period for the detection criterion to be valid.}, the values of the index are expected to be either $-1$ (saddle fixed point in the cell), $+1$ (center present), or $0$ (no fixed point).

\vspace{0.1cm}\noindent
\textbf{Fixed point extraction:} If the index is non-zero, find the location of the fixed point, which must satisfy $\bm{\Delta}_{\bar{p}} = \mathbf{0}$, using a Newton method. If a saddle-type fixed point is found, compute the eigenvectors of the monodromy matrix $\mathcal{M}$ (the spatial derivative of the flow map, \autoref{sec:theory}).

\vspace{0.1cm}\noindent
\textbf{Separatrices construction:} For each saddle-type fixed point, compute the stable and unstable manifolds (\ie separatrices) associated with its eigenvectors using an iterative shooting method~\cite{England:2005:Computing}.
\vspace{0.1cm}
\vspace{0.2cm}
\noindent
The existing method~\cite{Schlei:2014:Enhanced} suffers from significant limitations that make it unsuitable for mission design applications. Indeed, it can only find a very limited number of periodic orbits and corresponding fixed points, and does so at great computational expense. Specifically, the results reported in the paper~\cite{Schlei:2014:Enhanced} for the Earth-Moon system correspond to around 50 periodic orbits detected, a result that was only possible thanks to the tedious repeated application of the method in 100 small subdomains, which was required to use the index to detect fixed points in cells.
Similarly, the manifold construction method used previously was only successful in a small number of cases (\eg those related to Lyapunov orbits, see \autoref{sec:theory} and \autoref{fig:cr3bp_explanation}) because it is oblivious to the fundamental issue of \textit{transversality violation} that is discussed below. In contrast, our solution can reliably and more efficiently characterize the topological structure of the \pcrtbp to support an effective visual analysis.
For comparison, our method found 1450 periodic orbits in the same system by inspecting 2 large domains and 2 densely populated smaller domains around the Moon. Similarly, the manifold extraction algorithm presented hereafter succeeded in constructing most of the invariant manifolds corresponding to periodic orbits with limited instability ($|\nu_{SI}| < 10^5$, see \autoref{eq:SI}). Indeed, spacecrafts can maintain their position even on unstable orbits with manageable control effort, up to a certain value of the stability index. As previously stated, these topological objects are extremely valuable for our application as they can be used as anchor points and pathways through the dynamics of the \pcrtbp to chart a trajectory that satisfies mission objectives.

\subsection{Adaptive \Poincare map sampling}\label{sec:pmateC}
To permit a reliable detection of fixed points, the \Poincare index of the displacement map $\bm{\Delta}_p$ (\autoref{eq:delta}) must be evaluated around areas of the map that are small enough to contain at most a single fixed point for the considered period~\cite{Strogatz:1994:Nonlinear}. Hence, a very high resolution sampling yielding tiny cells is typically desirable. However, this approach, which was used in prior work~\cite{Schlei:2014:Enhanced}, is computationally prohibitive and a more subtle data-driven sampling is needed.

Our key observation is that the winding numbers associated with each trajectory (\autoref{sec:theory}) are locally smoothly varying characteristic parameters within regions of regular dynamics. Therefore, the values of all the winding number vectors $\mathbf{W}=(w_{xy}, w_{x\dot{x}}, w_{\dot{x}y})$ (\autoref{sec:existing approach}) measured in and around a cell should lie within a small interval. Hence, we apply this criterion in each cell to the $\mathbf{W}$ values measured at the vertices of the cell, as well as to the trajectories that intersected the \emph{interior} of the cell during the \Poincare map sampling. If this criterion is not satisfied, the cell is marked for subdivision.

Practically, we use a quadtree-like data structure that records winding numbers both at the sampling vertices and inside the cells. Since all \Poincare section crossings of a given trajectory share the same winding numbers set $\mathbf{W}$ as the initial vertex, each one of the $p$ intermediate returns is assigned the same $\mathbf{W}$ values. These values are added to the cell containing the return and are then tested as part of the subdivision criterion discussed previously. When required, cells are subdivided with internal data assigned to the corresponding quadrant within the original cell.
A user-specified maximum depth level parameter, $d_{max}$, is employed to represent the total number of subdivision layers allowed. Cells are also subdivided if any corner has an undefined winding number, which occurs in chaotic regions.

\begin{figure}[hbt]
 \centering
 \includegraphics[width=0.99\columnwidth]{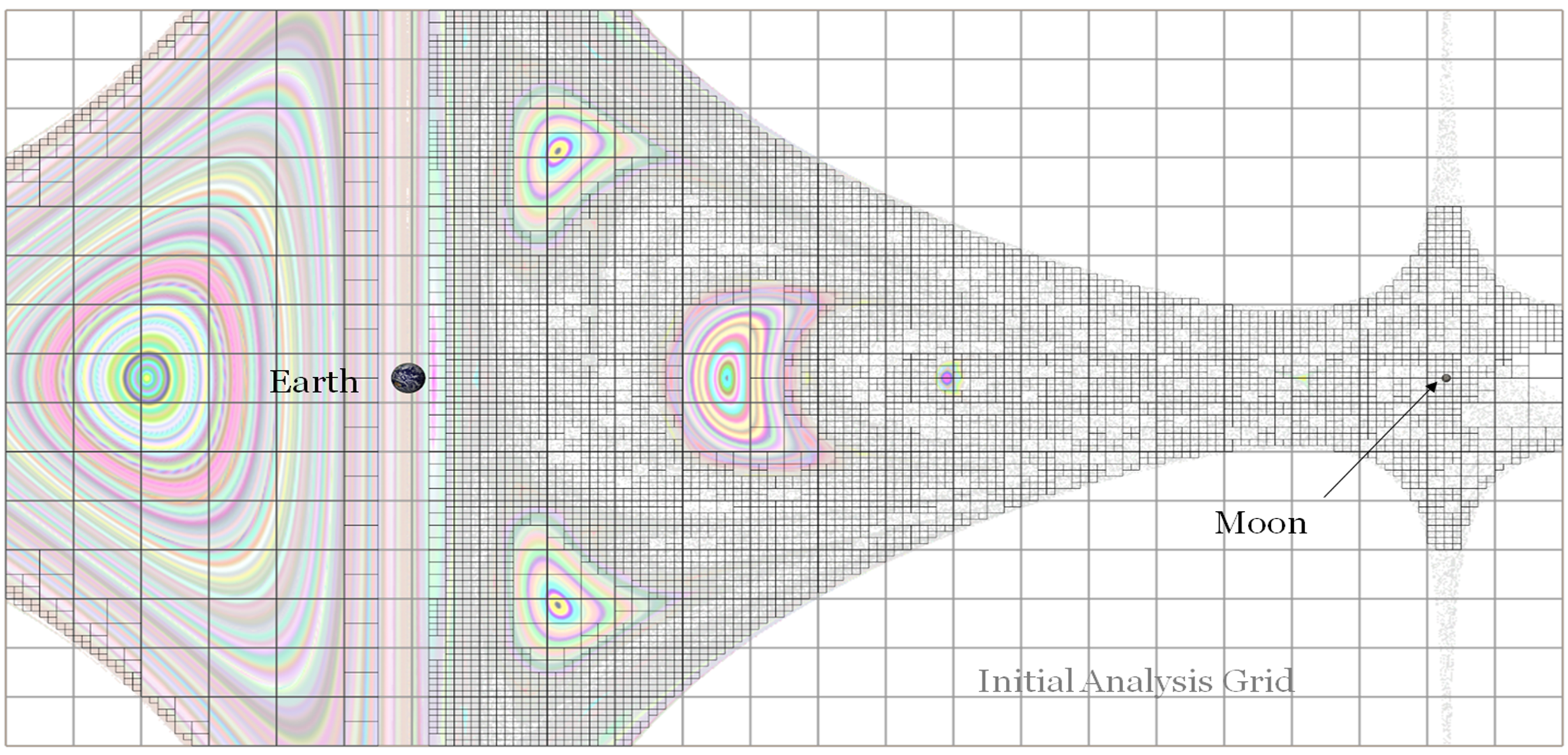}
 \caption{Adaptive cell subdivision based on the winding number set $\mathbf{W}$ applied to the domain $D_{EM}$ with parameters $C=2.96$ and $d_{max}=3$.
  }
 \label{fig:AdGrid}
\end{figure}

\autoref{fig:AdGrid} shows the adaptive resolution mesh produced in the domain $D_{EM}$ for a maximum refinement depth $d = 3$. The initial grid is shown in thick gray lines. The highest resolution is achieved in chaotic regions where the dynamics is most complex. This result is excellent from an astrodynamics perspective since the saddles embedded in chaos offer the most versatile transfer opportunities. Regions of regular dynamics, in contrast, are coarsely resolved, as expected.

\subsection{\Poincare section transversality}\label{sec:transversality}
\begin{figure}[hbt]
    \centering
	\label{sf:secTan}
	\includegraphics[width=0.40\columnwidth]{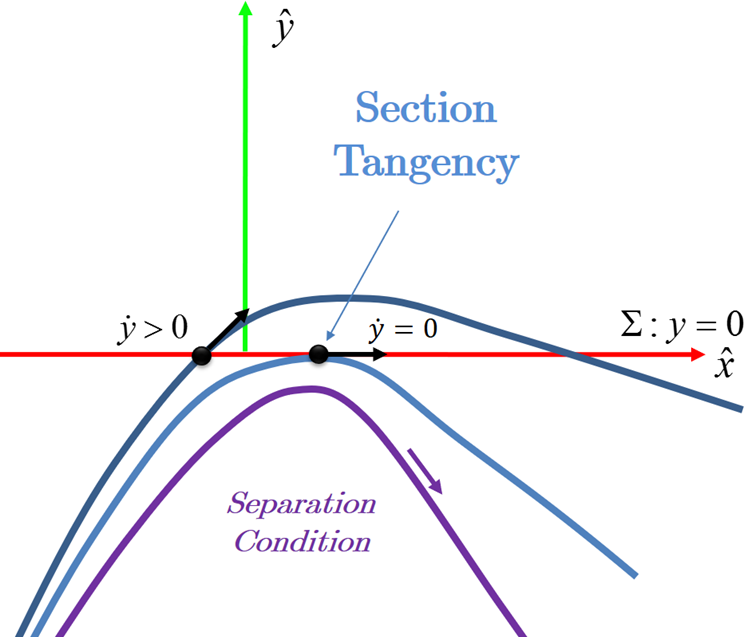}
	\label{sf:singularity}
	\includegraphics[width=0.58\columnwidth]{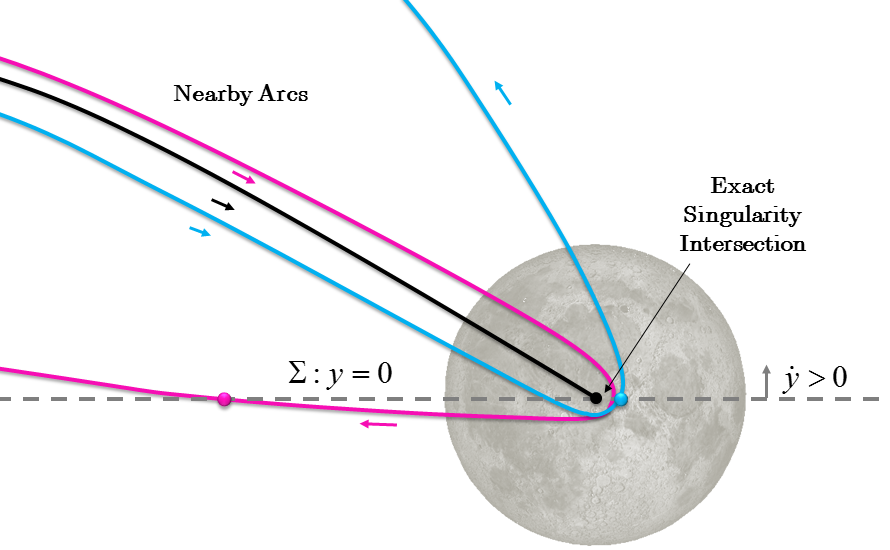}
 \caption{Transversality violation types in the \pcrtbpNoSpace. Left: section tangency, Right: singularity intersection.}
 \label{fig:mapTV}
\end{figure}

An issue that can prevent the computation of the \Poincare index in prior work~\cite{Schlei:2014:Enhanced} is the presence of discontinuities of the $\bm{\Delta}$ mapping along cell edges.
Two properties of the \pcrtbp can explain this behavior: highly sensitive dynamics and transversality violation of the flow map for the chosen section $\varSigma$. Indeed, the the ability of the \Poincare map to capture the dynamics of the system is premised on the assumption that the flow is \emph{transverse} to the chosen section in the considered region.
Transversality violations are typically the result of one of two specific trajectory events.  First, trajectories that are tangent to the section along their path generate discontinuities in $\bm{\Delta}$, see \autoref{fig:mapTV}.
A second event is an intersection by the trajectory of a primary body in \crtbp (see \autoref{fig:mapTV}).

\subsection{Resolving the \Poincare Index}\label{ssec:resolvePI}
We perform the evaluation of the \Poincare index in cells where transversality violations occur by considering the behavior of $\bm{\Delta}$ (\autoref{eq:delta}) in the limit approaching a transversality violation.
A discontinuity of $\bm{\Delta}$ at some location $\mathbf{g}$
on the closed curve $\Gamma$ (\ie{} the boundary of a cell) for period $p$ creates a discontinuity in the angle coordinate of $\bm{\Delta}$ at $\mathbf{g}$ (\autoref{fig:poincare_index}, D.). However, since the limits of the value $\bm{\Delta}(\mathbf{g})$ exist on both sides of $\mathbf{g}$, the \Poincare index can be expressed as the summation of improper integrals
\begin{equation}\label{eq:index2}
 \kappa = \frac{1}{2\pi}\oint_\Gamma d\alpha(\bm{\Delta}) = \frac{1}{2\pi}\left( \int_{\mathbf{\gamma}_0}^{\mathbf{g}} d\alpha(\bm{\Delta})
+\int^{\mathbf{\gamma}_0}_{\mathbf{g}} d\alpha(\bm{\Delta})\right),
\end{equation}
where $\mathbf{\gamma}_0$ is a starting point along $\Gamma$ ($\mathbf{\gamma}_0\neq\mathbf{g}$) and $\alpha(\bm{\Delta})$ is the angle formed by $\bm{\Delta}$ with the $\mathbf{x}$ axis.

The edge sampling approach used in prior work~\cite{Schlei:2014:Enhanced} is therefore modified to detect transversality violations and ensure fine sampling of the edge around the corresponding discontinuity.

\subsection{Fixed Point Extraction}\label{ssec:pmateFPGuess}

The success of a numerical search depends heavily on the quality of the initial guess.
Our solution starts by sampling the displacement map $\bm{\Delta}$ at a set of regularly distributed positions within the cell with nonzero index.
Practically, if the considered variable is $\mathbf{\zeta} = \mathbf{x} - \mathbf{s}$
with $\mathbf{s}$ representing the (unknown) saddle-type fixed point location and $\mathbf{x}=(x,y)$ is the dynamic state of the system (see \autoref{eq:CR3BP}),
then a \emph{quadratic} model of the sampled dynamics is formed as
\begin{equation}\label{eq:SaddleModel}
 \mathbf{\dot{\zeta}} = A_s\mathbf{\zeta} + \tfrac{1}{2}\mathbf{\zeta}^T \underline{Q}\mathbf{\zeta}.
\end{equation}
Note, $A_s$ is a $2\times2$ matrix, and $\underline{Q}$ is a $2\times2\times2$ tensor comprised of the coefficients of nonlinear terms in the model, with $\underline{Q}=0$ in the linear model.  A Levenberg-Marquardt optimization process~\cite{Press:2007:Numerical} is then applied to fit the model to the sampled data and infer the approximate location of the fixed point $\mathbf{\zeta} = \mathbf{0}$, which yields an initial guess for an iterative numerical search.

The Newton search and multiple shooting\footnote{Multiple shooting methods introduce intermediate steps to the solution.} solutions used in prior work to find fixed points often fail. To remedy this situation, we adopt a multipronged approach.
Starting with a single shooting method for fixed point refinement, we switch to a multiple shooting method if this first attempt fails. If the second solution fails as well, we apply a refined precision Newton method, which is significantly more computationally expensive than the previous two but possesses stronger convergence properties.

\subsection{Invariant Manifold Extraction}\label{sec:manifold}
If a saddle type is identified, the construction of the stable and unstable manifolds constitutes the last step of our topology extraction and we derive eigenvectors and stability index (\autoref{eq:SI}) from $\mathcal{M}$.
Prior work~\cite{Tricoche:2011:Visualization,Schlei:2014:Enhanced} constructs invariant manifolds through a series of two-point boundary problems that aim to ensure smoothness and fine sampling of the manifold~\cite{England:2005:Computing}.
Unfortunately, this solution does not handle the issue raised by transversality violations in the \pcrtbp (see \autoref{sec:transversality}), nor does it provide any guidance to accommodate the numerical challenges associated with this particular system. We describe in the following our improvements of this method. We then present a data structure that allows the mission designer to instantly navigate the manifolds up- and downstream as demonstrated in \autoref{sec:results}.

\begin{figure}[hbt]
   \centering
   \includegraphics[width=0.99\columnwidth]{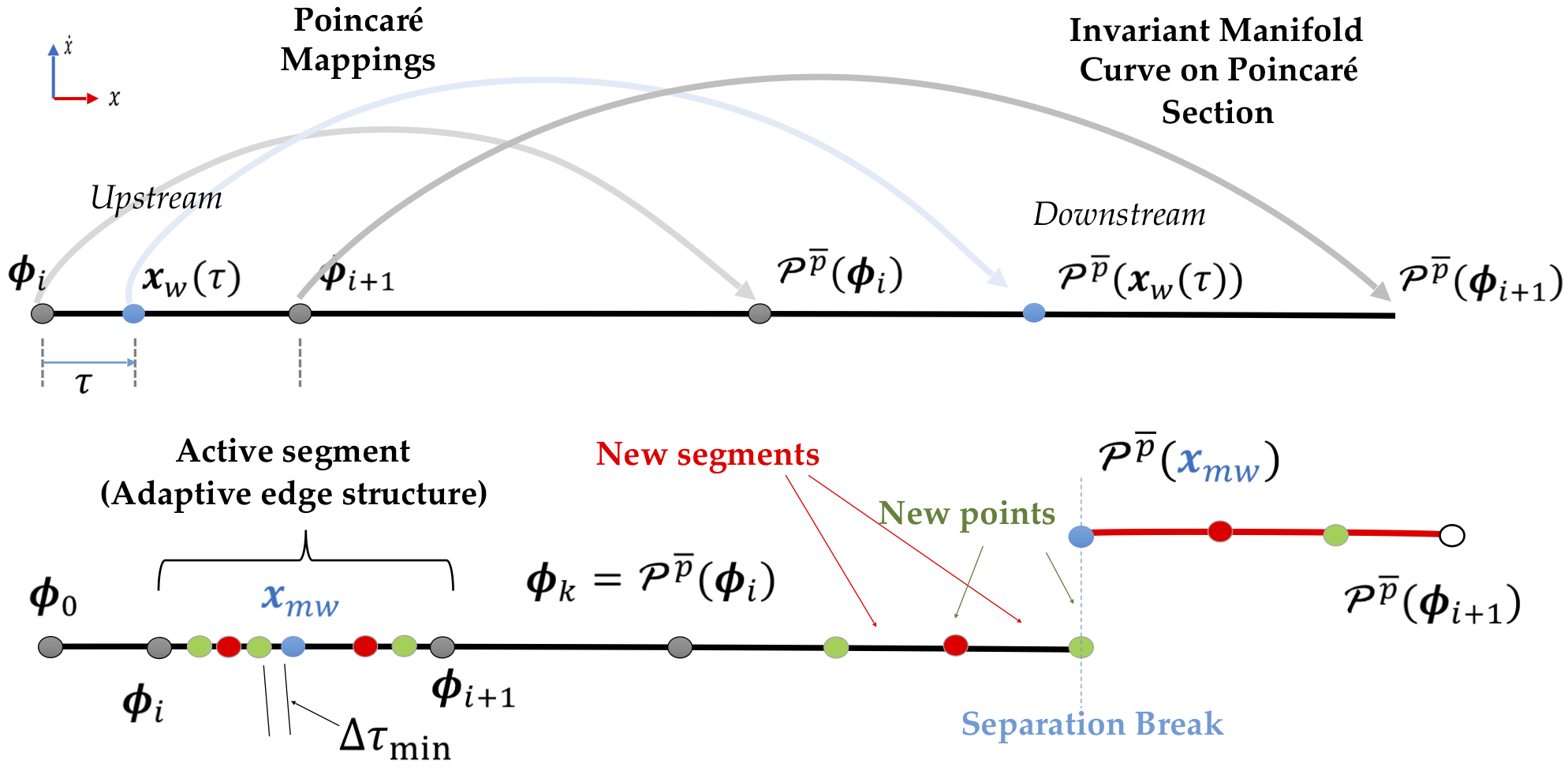}
   \caption{Top: Schematic of a 1D invariant manifold curve on the \Poincare section. Bottom: Generating new downstream manifold points and segments through a transversality violation.}
   \label{fig:manDef}\label{fig:mAlgoTVSeg}
\end{figure}

\vspace{0.5em}
\noindent\textbf{Manifold Extraction with Curve-Refinement.}
Consider two adjacent positions $\mathbf{\phi}_i$ and $\mathbf{\phi}_{i+1}$ that form a segment $w=\overline{\mathbf{\phi}_i\mathbf{\phi}_{i+1}}$ on the manifold. We further assume that both positions are close enough such that linear interpolation between these two positions yields positions that are themselves on the manifold, which is a valid approximation as long as the manifold does not come too close to another saddle fixed point. Applying the \Poincare map $\mathbf{\mathit{P}}^{\bar{p}}$ ($\bar{p}$ is the period of the corresponding saddle fixed point) to any such intermediate position will result in a new position further downstream on the manifold. Refer to \autoref{fig:manDef} (top).
The algorithm used in prior work~\cite{Schlei:2014:Enhanced} performs an adaptive sampling of the segment $w$ by the \Poincare map, controlled by curve quality checks, to construct the next segment $=\overline{\mathbf{\phi}_{i+1}\mathbf{\phi}_{i+2}}$ on the manifold and ultimately extract the entire manifold~\cite{England:2005:Computing}.
Our solution follows the same approach while simultaneously checking for transversality violations.

The heuristics for detecting transversality violations during \Poincare index evaluation are reapplied alongside the curve-refinement criteria.  If a downstream transversality violation is detected between consecutive segment samples, the segment is bisected on that interval.  Subdivision continues until the distance between consecutive points reaches a user-prescribed minimal distance ($u_{\min}$, which corresponds to a relative distance $\bm{\Delta}\tau_{\min}$ on the segment).
An example is depicted in \autoref{fig:mAlgoTVSeg} (bottom) where downstream mappings are color-coded by their initial position on the active segment: a downstream transversality violation exists between $\mathbf{\phi}_i$ and the midpoint $\mathbf{x}_{mw}$, and subdivision localizes the separation condition when the parameter differential is below $\bm{\Delta}\tau_{\min}$.

\vspace{0.5em}
\noindent\textbf{Stopping Criteria.}\label{ssec:manStop}
We found two criteria primarily effective in controlling the useful downstream length of an invariant manifold as it starts to form tangles in the vicinity of another saddle point. Refer to \autoref{fig:topology}. The first stopping criteria tracks a practical measure for spacecraft trajectory planning. The manifold construction algorithm creates a parent-child link between upstream and downstream segments. Our algorithm caps the manifold progression by stopping when the depth of the resulting tree reaches a maximum depth $d_{w,\max} = 5$ (or $\mathbf{\mathit{P}}^{5\overline{p}}(\mathbf{x})$).
The second stopping condition observes simultaneous advection of manifolds from the same periodic orbit for the detection of so-called \emph{homoclinic connections}.

\vspace{0.5em}
\noindent\textbf{Screening Computations}\label{sec:screening}
Given the high computational cost of manifold construction, we perform several tests beforehand to prevent unnecessary computations. First, we pre-screen for potentially impractical structures for spaceflight, \ie periodic orbits with exceptionally high instabilities.  A threshold cutoff is established on stability index magnitude at $|\nu_{SI}|>10^6$ (refer to \autoref{eq:SI}), whereby a lower $|\nu_{SI}|$ cutoff can be used to further reduce the overall workload of our method.

A second important observation is that a lower bound is necessary for $u_{\min}$ (which regulates upstream manifold segment subdivisions) for realistic spaceflight. The accuracy limitations of sensors and engines prevent an exact knowledge of the spacecraft position and speed and make maneuvers impractical below a certain \dv magnitude.
Spacecraft state determination outside of low Earth orbits is limited to an accuracy of 3 km for position and 0.1 mm/s for velocity based on measurement error of standard capabilities~\cite{Yim:2000:Autonomous}. Practically, we require $u_{\min}$ values above $2\times10^{-5}$ (nondimensional map displacement) for the Earth-Moon system, which is equivalent to 2.05 cm/s for velocity and 7.69 km for position. Note that the value of
$u_{\min}$ is different across \crtbp systems. Suggested values are provided in appendix.
\ignore{
Refer to \autoref{tab:manifoldParams} for suggested values.

\begin{table}[ht]
   \centering
   \resizebox{\columnwidth}{!}{%
   \begin{tabular}{|l|c c c c c c|}
     \hline
     System & $u_{\min}$ & $\bm{\Delta}_{\min}$ & $\bm{\Delta}_{\max}$ & $\alpha_{\max}$ & $(\bm{\Delta}\alpha)_{\max}$ &  $\lvert\nu_{SI}\rvert_{\max}$ \\
     \hline \hline
     EM      & $2\times10^{-5}$ & $1\times10^{-5}$  & 0.1 & 0.3 ($17.2^\circ$) & 0.001 & $2.5\times10^3$ \\
     SEnc & $1\times10^{-6}$ & $1\times10^{-6}$ & 0.05 & 0.1 ($5.7^\circ$) & 0.001  & $5.0\times10^3$ \\
     \hline
    \end{tabular}}
    \caption{Heuristic parameters employed for invariant manifold advection in the indicated \crtbp systems. Phase space displacement values are listed in nondimensional units (as defined by England \etal\cite{England:2005:Computing}).}
    \label{tab:manifoldParams}
\end{table}
}

\vspace{0.5em}
\noindent\textbf{Manifold Arc Extraction} Implementing manifolds within trajectory design applications requires the ability to select an invariant manifold state from the \Poincare map visualization and reconstitute the arc. Simply propagating a user-selected invariant manifold state upstream (\ie{} forward-time for $W^S$ and reverse-time for $W^U$) often fails because of the flow-dividing nature of the invariant manifolds, from which neighboring trajectories diverge exponentially.

\begin{figure}[ht]
 \centering
 \includegraphics[width=\columnwidth]{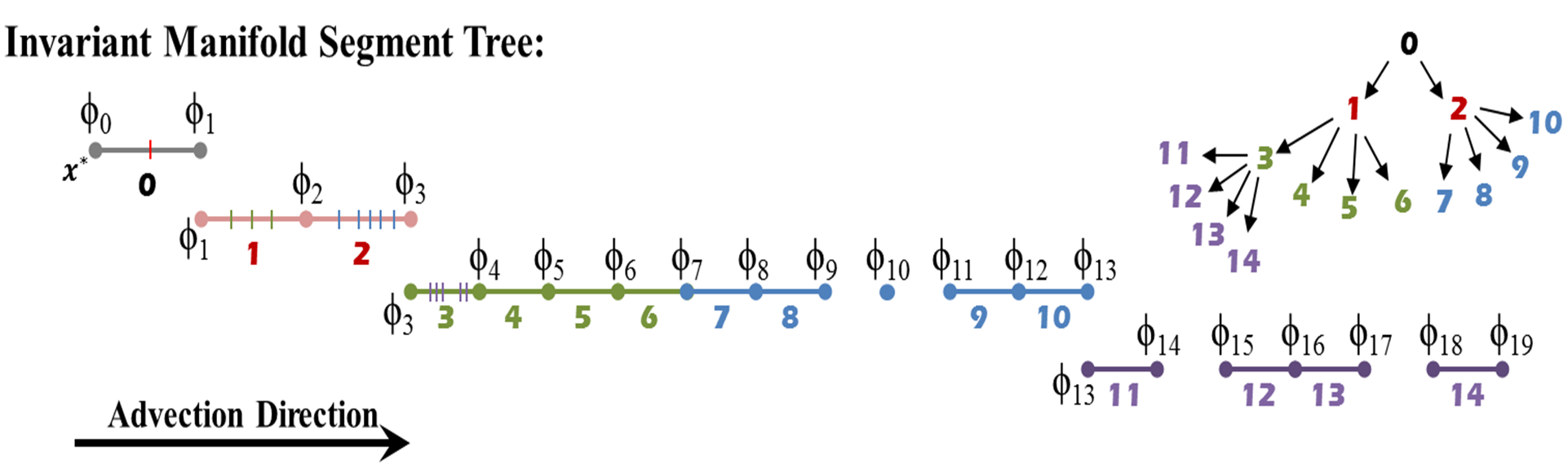}
 \caption{As the construction of the invariant manifold progresses via curve-refinement, the spawning of new manifold segments generates a tree structure that can be employed for accessing data.}
 \label{fig:manTree}
\end{figure}

Upstream manifold arc reconstruction with our method circumvents the challenges associated with propagating manifold arcs upstream with a very effective data structure format resulting from the manifold advection procedure.  Our manifold extraction process advects states sampled from an upstream manifold segment to create a new group of downstream segments through the \Poincare map; this procedure naturally forms a \textit{manifold segment tree}.
To illustrate, sample manifold segments near the origin location of the advection procedure are arranged as a staircase schematic indicating depth levels ($d_w$) as in \autoref{fig:manTree}; each step down symbolizes the downstream progression to the next group of segments after $\overline{p}$ map iterates.
To reconstruct the entire upstream trajectory from the fixed point to an arbitrary user-selected point, the downstream mappings recorded during manifold construction
supply a reliable framework to return the intermediate trajectory states between the parent and child manifold segments. Practically, the local (linear) coordinates of a state on a child segment are mapped upstream to a state with the same local coordinates on the parent segment, which eventually leads to the immediate vicinity of the saddle fixed point from which the manifold originated.

\section{Topology Visualization}\label{sec:results}

We present in this section the visualizations of the \pcrtbp that were made possible by our topology extraction method and illustrate the variety of structures that can be observed in this context.

\subsection{Fixed Points Discovery}\label{sec:OrbitResults}

We visualize in the following the fixed points that our method was able to discover. Among these fixed points are many that are either challenging to extract with existing means (\eg{} they require very high resolution sampling in a specific region and lengthy numerical search), or are presented here for the first time. It is important to note that the topological complexity (including the number of fixed points\footnote{We only consider fixed points of period $p<p_{max}$}) of the \pcrtbp varies significantly with the considered energy level, as quantified by the Jacobi constant $C$ (\autoref{eq:jc}). Specifically, high values of $C$ correspond to low energy and exhibit a simpler topology than lower $C$ values, \ie{} higher energy levels. Both cases are considered next to illustrate this effect.

\begin{figure}[hbt]
 \centering \includegraphics[width=\linewidth]{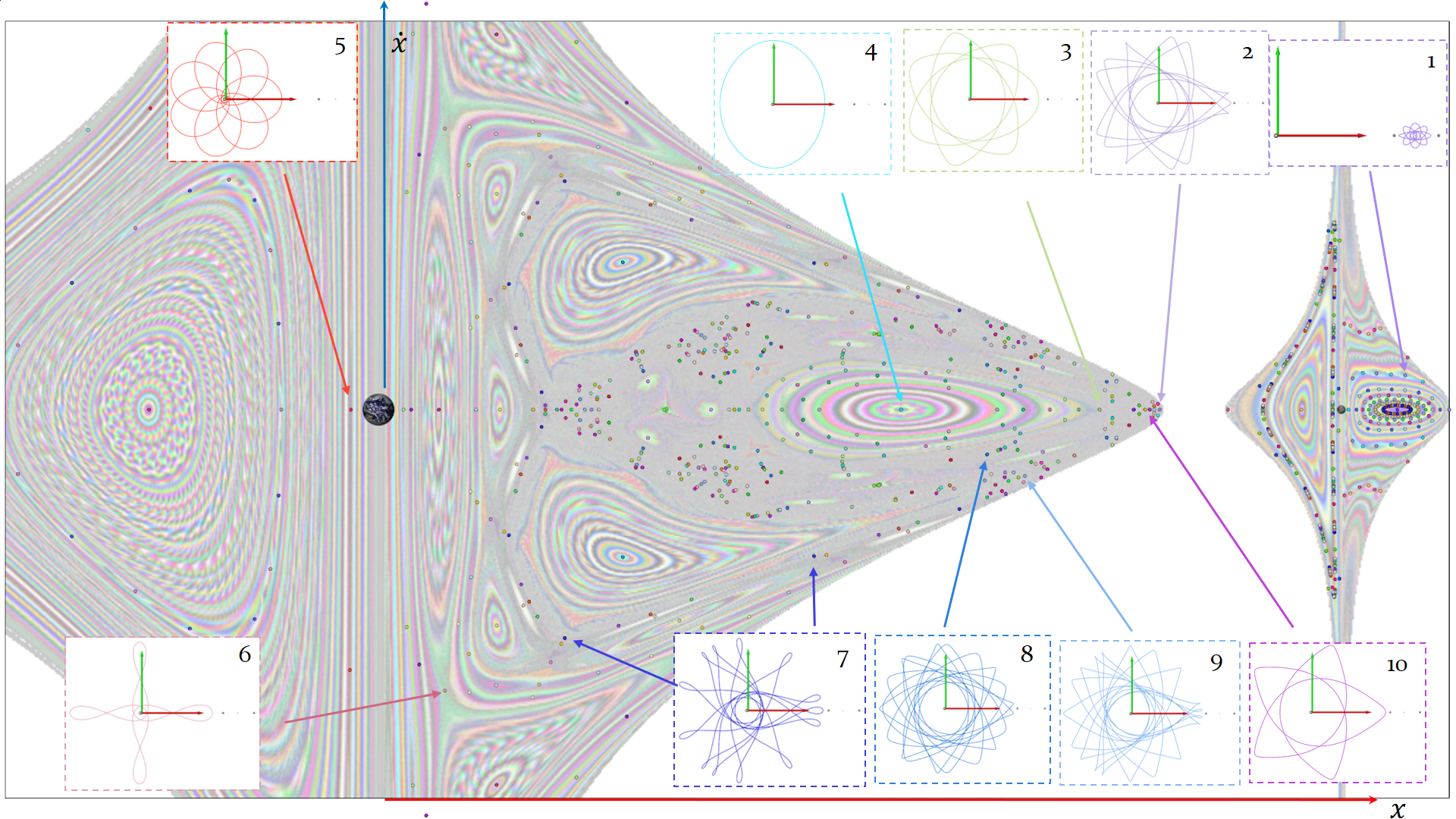}
 \caption{Fixed points and selected periodic orbits found in the Earth-Moon system at $C=3.2$.}
 \label{fig:emfpC3_20}
\end{figure}

\label{ssec:fpem}
Practically, in the EM system, $C=3.2$ is a low energy level\footnote{This energy level is still much higher than typical levels in low Earth orbit, for which $C \in \left[ 8, 16\right]$} in which it is impossible to reach the Moon from the Earth as they belong to disconnected regions of the dynamics. In this case, our visualization of the identified fixed points, shown in \autoref{fig:emfpC3_20}, reveals
saddles and centers grouped in island chains (repeated patterns similar to what is shown in \autoref{fig:topology}), as well as isolated
saddle-type fixed points embedded in chaotic regions. In this and following images, the fixed points belonging to the same periodic orbit are assigned the same color. Nonetheless, the large number of periodic orbits makes it challenging to visually identify all the fixed points associated with any given orbit. It is instead more insightful to analyze the geometry of specific periodic orbits while highlighting their associated fixed points. Their visualization, displayed in the $xy$ plane in \autoref{fig:emfpC3_20}, shows each selected periodic orbit as a closed curve, sometime with confounding geometric complexity. The background image is formed by orbital convolution, an extension of line integral convolution method~\cite{Cabral:1993:Imaging} to maps~\cite{Tricoche:2011:Visualization,Schlei:2014:Enhanced}. At this energy level, transversality violations are rare within the considered region and the fixed point extraction is relatively straightforward with little need for cell subdivision.

A more complex case is considered next, in which a higher energy level is selected where chaos exists throughout the domain of interest. At $C=2.96$ the Earth-Moon system permits trajectories everywhere in the $xy$ plane, and in particular between the Earth and the Moon.
Broader sampling parameters are applied over a larger analysis domain (see Trial 1 in \autoref{tab:pmateEM}). We find fixed points throughout $D_{EM}$ but only a small number in close proximity of the Moon. Increasing the sampling resolution near the Moon allows us to uncover a large number of additional fixed points in that region.
Refer to Trials 2, 3, and 4 in \autoref{tab:pmateEM}.  As shown in \autoref{sf:emfpFull}, many fixed points were extracted for $C=2.96$ in the EM system, reaching a total of 1450 distinct periodic orbits. A closeup view centered on $L_1$ reveals the density of fixed points in that region.

\begin{table}[ht]
  \centering
  {\small
  \begin{tabular}{ c c | c | c c c}
  Trial & $C$ & Domain $(x, \dot{x})$ (nondim) & Resolution & $l_{\min}$ & $p_{\max}$ \\
  \hline
  0 & 3.2 & $[0.4,1.1] \times [-2.5,2.5]$  & $24\times16$ & 8\e{-5} & 12 \\
  1 & 2.96 & $[0.4,1.1] \times [-2.5,2.5] $ & $24\times16$ & 8\e{-5} & 12 \\
  2 & 2.96 & $[0.9,1.0] \times [-1.5,1.5]$ & $8\times8$ & 2\e{-5} & 12 \\
  3 & 2.96 & $[0.78,0.92] \times [-0.4,0.4]$ & $8\times8$ & 2\e{-5} & 6 \\
  4 & 2.96 & $[0.9925,1.08] \times [-0.2,0.2]\rbrace$ & $6\times6$ & 2\e{-5} & 4 \\
  \hline
  \end{tabular} }
  \caption{Parameters used in the Earth-Moon system.}
  \label{tab:pmateEM}
\end{table}

Among the fixed points found at $C=2.96$, many novel saddle-type periodic orbits were identified. As shown in \autoref{sf:emfpFull}, several periodic orbits are commonly known such as Orbit 4 (the $L_1$ Lyapunov), Orbit 3 (stable 3:2 resonant orbit, which means that it completes 3 revolutions around the Earth while the Moon completes two), and Orbit 5 (the $p=3$ unstable DRO - quasi-periodic island near the Moon).  Several orbits, though, transit between the interior and exterior regions (such as Orbit 2) and DRO vicinity to exterior or interior (\eg{} Orbit 1).  Yet others like Orbit 6 visit all the aforementioned regions, making such orbits compelling for transfer design.
Though the analysis is only performed within the primary analysis domain $D_{EM}$ on the $\varSigma: y=0$ \Poincare section (as per \autoref{tab:pmateEM}), many unstable periodic orbits that cross this section travel well beyond it, to $L_3$, $L_4$, and $L_5$ vicinities.  Clearly, our results offer a vivid dynamical understanding of this particular system.

\subsection{\Poincare Map Topology Structure}\label{sec:skeleton}

\begin{figure}[hbt]
 \centering
 \subfigure[Primary analysis domain $D_{EM}$]{\label{sf:emTSC3_2}\includegraphics[width=\linewidth]{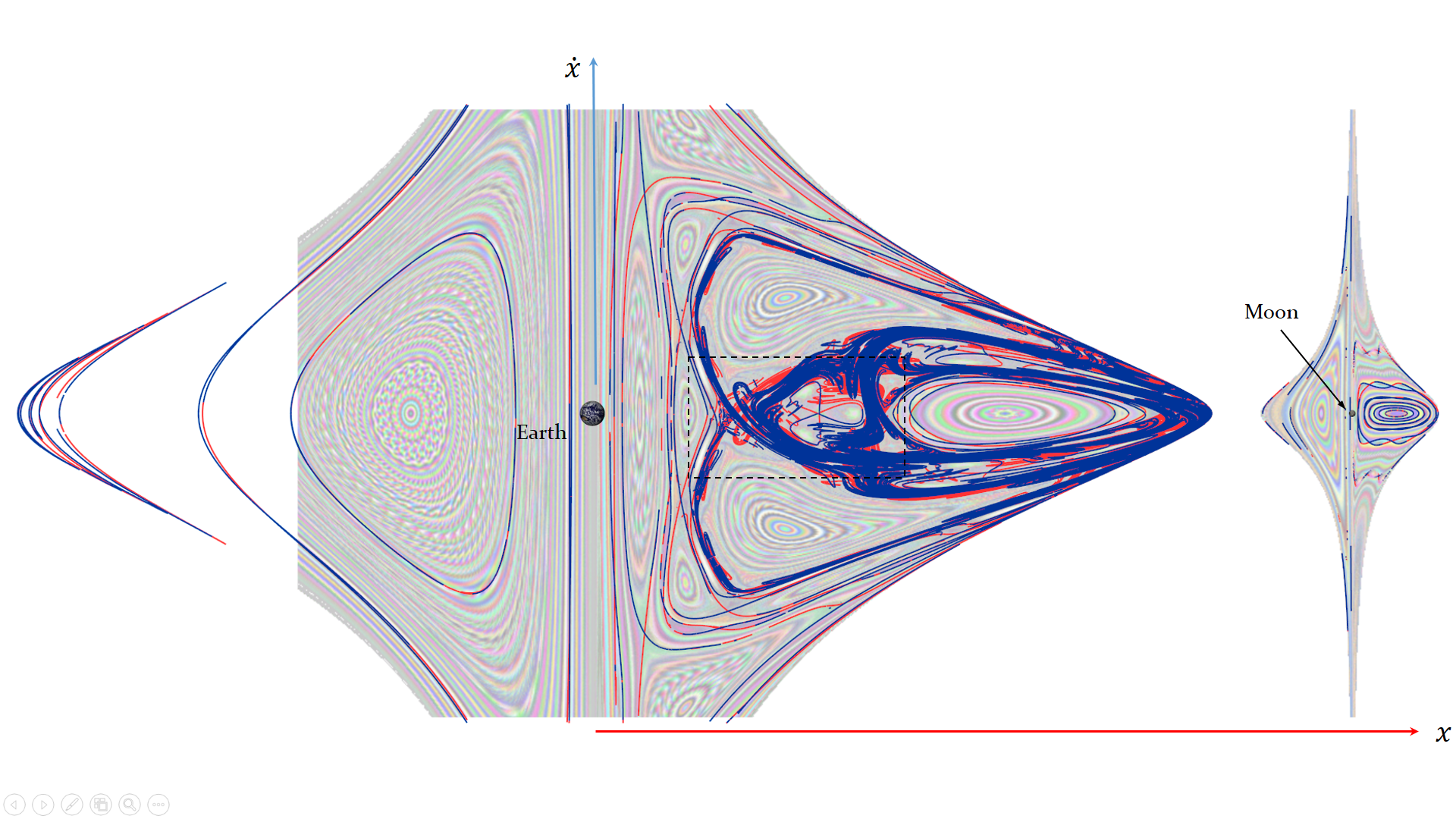}}
\subfigure[Zoom-in on indicated domain]{\label{sf:emtsC3_2Zoom}\includegraphics[width=\linewidth]{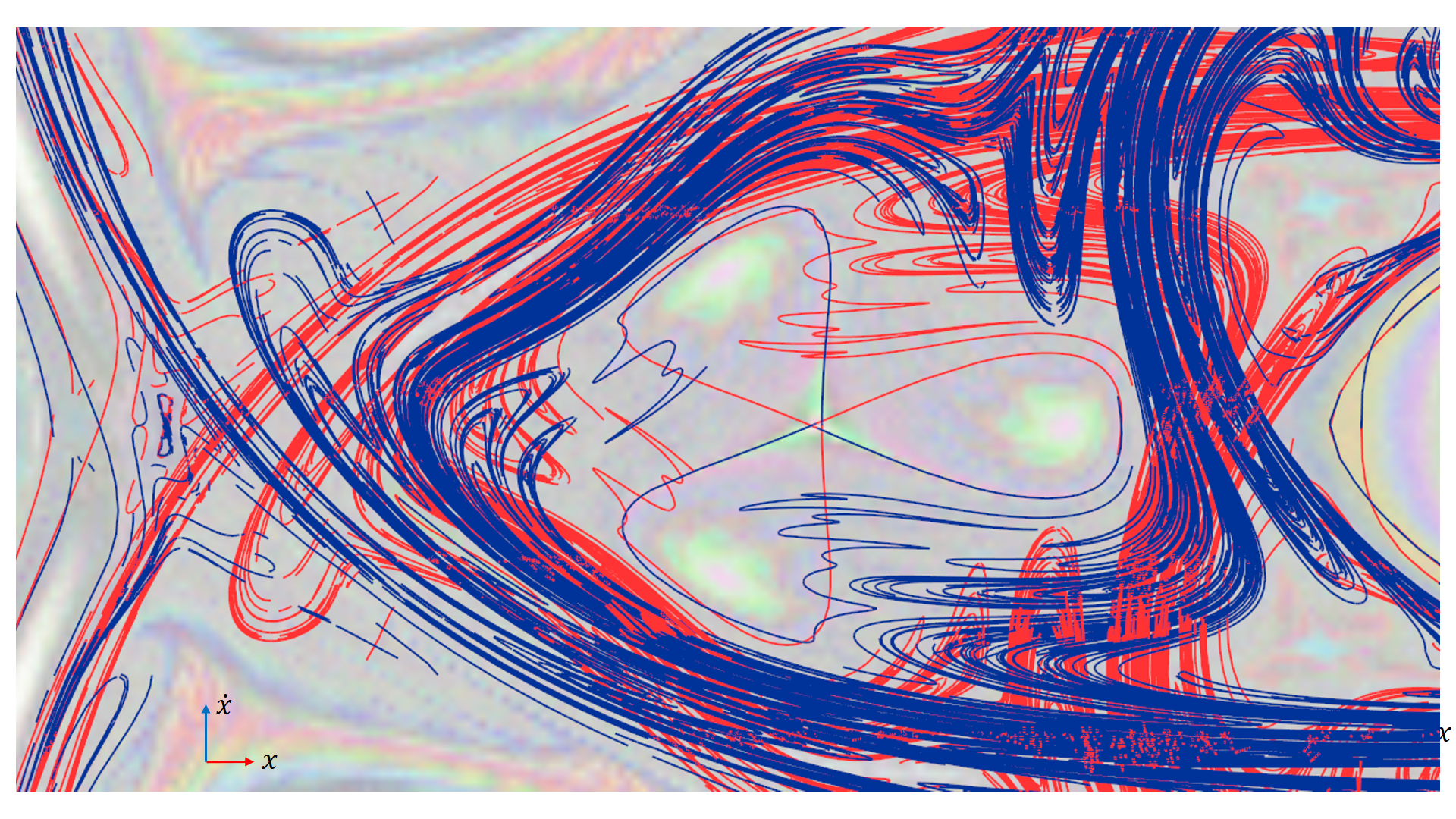}}
 \caption{The \Poincare map topology skeleton (unstable manifolds in red and stable in blue) computed with the manifold extraction algorithm in the Earth-Moon system within the domain $D_{EM}$ at $C=3.2$.}
 \label{fig:emTSC3_2}
\end{figure}

With fixed points extracted, the \Poincare map topology structure is further characterized by means of our invariant manifold extraction algorithm, which
is first demonstrated in the Earth-Moon system at $C=3.2$.  The large-scale topology extraction result appears in \autoref{fig:emTSC3_2} with unstable manifolds ($W^U$) and stable manifolds ($W^S$) colored in red and blue, respectively. At $C=3.2$, invariant manifolds are extracted throughout the chaotic areas, thoroughly filling in the phase space areas between quasi-periodic islands.  Our algorithm captures saddle-center island chains except on some islands near the Moon.  Difficulties near the Moon can be explained by numerical sensitivity and numerical error build-up during integration as trajectories pass exceptionally close to the singularity multiple times before completing the $\overline{p}$-th iterate. As with fixed point extraction, constructing invariant manifolds for the Earth-Moon system at $C=3.2$ is rather easy and only a small number of transversality violations are encountered.
In fact, this manifold set is processed without stability index pre-screening (see \autoref{sec:screening}) and still completes the advection procedure faster than for higher energy systems. Yet, visualizing the \Poincare map topology skeleton remains challenging, as can be seen in a close-up representation (\autoref{sf:emtsC3_2Zoom}).  Artifact segments shortcut tangles in both manifold types, which are due to loose curve-refinement parameters. Chaotic tangles, on the other hand, strongly influence the generation time and geometric complexity of invariant manifolds.

\begin{figure}[ht]
 \centering \includegraphics[width=0.99\columnwidth]{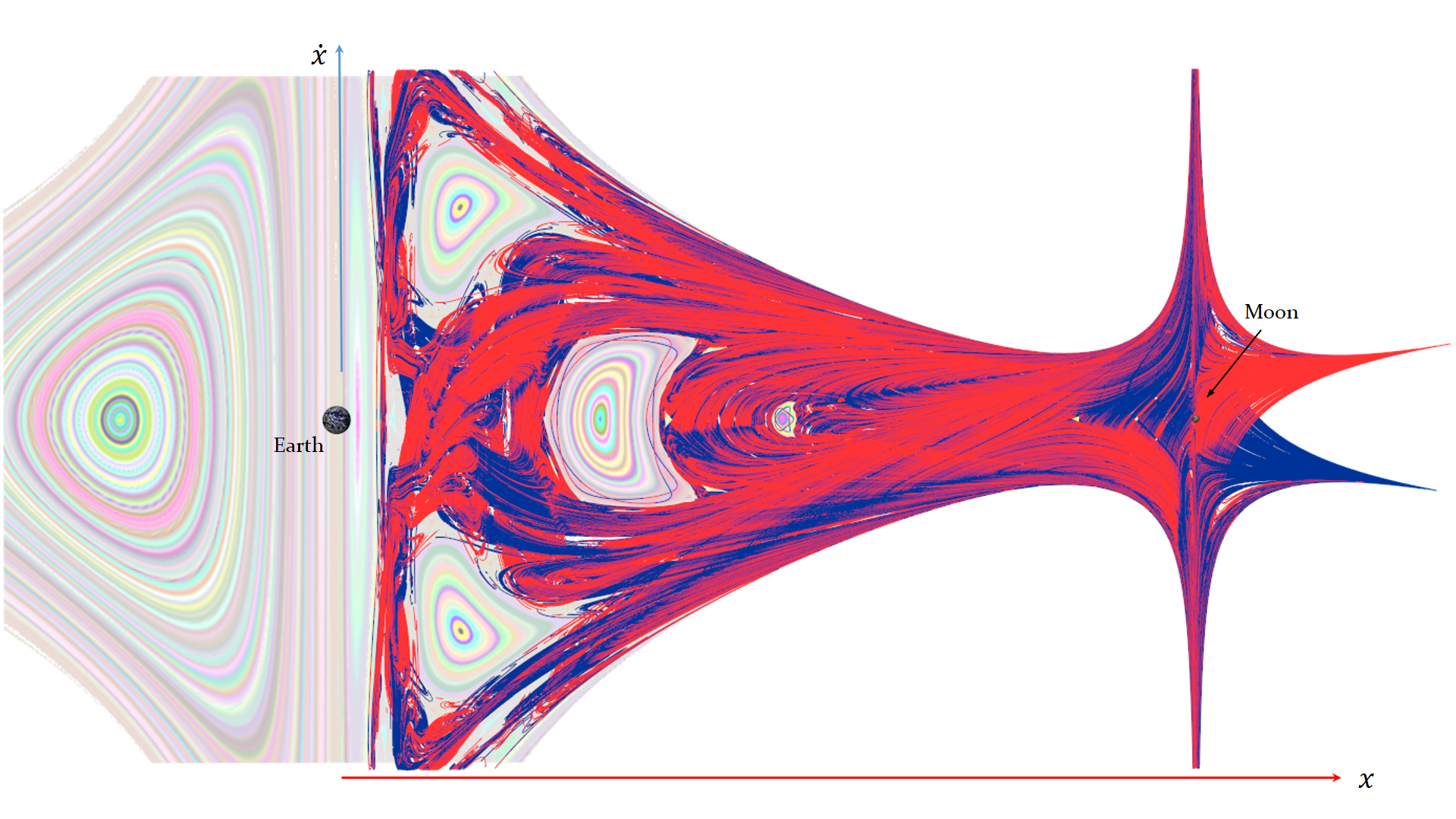}
 \caption{The \Poincare map topology skeleton (unstable manifolds in red and stable manifolds in blue) computed in the Earth-Moon system at $C=2.96$.}
 \label{sf:emTSC2_96}
\end{figure}

The manifold extraction method is next applied at a higher energy level.
At $C=2.96$ in the \textit{EM} system, our manifold construction produces a visualization of both stable (blue) and unstable (red) manifolds for the periodic orbits shown in \autoref{sf:emTSC2_96}. We apply stability filtering ($|\nu_{SI}|\leq 2500$) to mitigate the visual and structural complexity. Despite a severe reduction in the number of manifolds,  the set of manifolds densely populates almost the entire chaotic region.
As can be seen, the only areas within the chaotic region without stable manifolds correspond to trajectories that escape the entire Earth-Moon system.

\section{Application to Trajectory Design}\label{sec:application}
We present in the following the insights gained from the application of our topology visualization technique to two representative trajectory design problems. As shown below, the topological scaffold formed by periodic orbits and invariant manifolds, and further enhanced by interactive design support, broadens the design possibilities available to domain experts, delivering new options and the ability to quickly examine trade-off decisions.

\subsection{Topology-Assisted Visual Design}\label{sec:emTopo}\label{sec:design}

The manifold arc selection solution presented in \autoref{sec:manifold} was created to allow for the interactive visual inspection of transfer solutions along manifolds. Specifically, this selection mechanism allows the user, for any arbitrary location selected interactively along a manifold, to instantly determine the precise path (including entry point) that a spacecraft would need to follow to arrive at that location. In particular, given a position where two manifolds intersect in the visualization, the user can readily query the path, upward and downward, that will be traveled from that point with no or minimal correction need. This makes it possible to use the visualization of the topology to quickly examine transfer possibilities.

\begin{figure}[hbt]
 \centering
 \subfigure[Manifolds of the $L_1$ Lyapunov and $p=3$ DRO periodic orbits]{\label{sf:interLyapDROMap}\includegraphics[width=\linewidth]{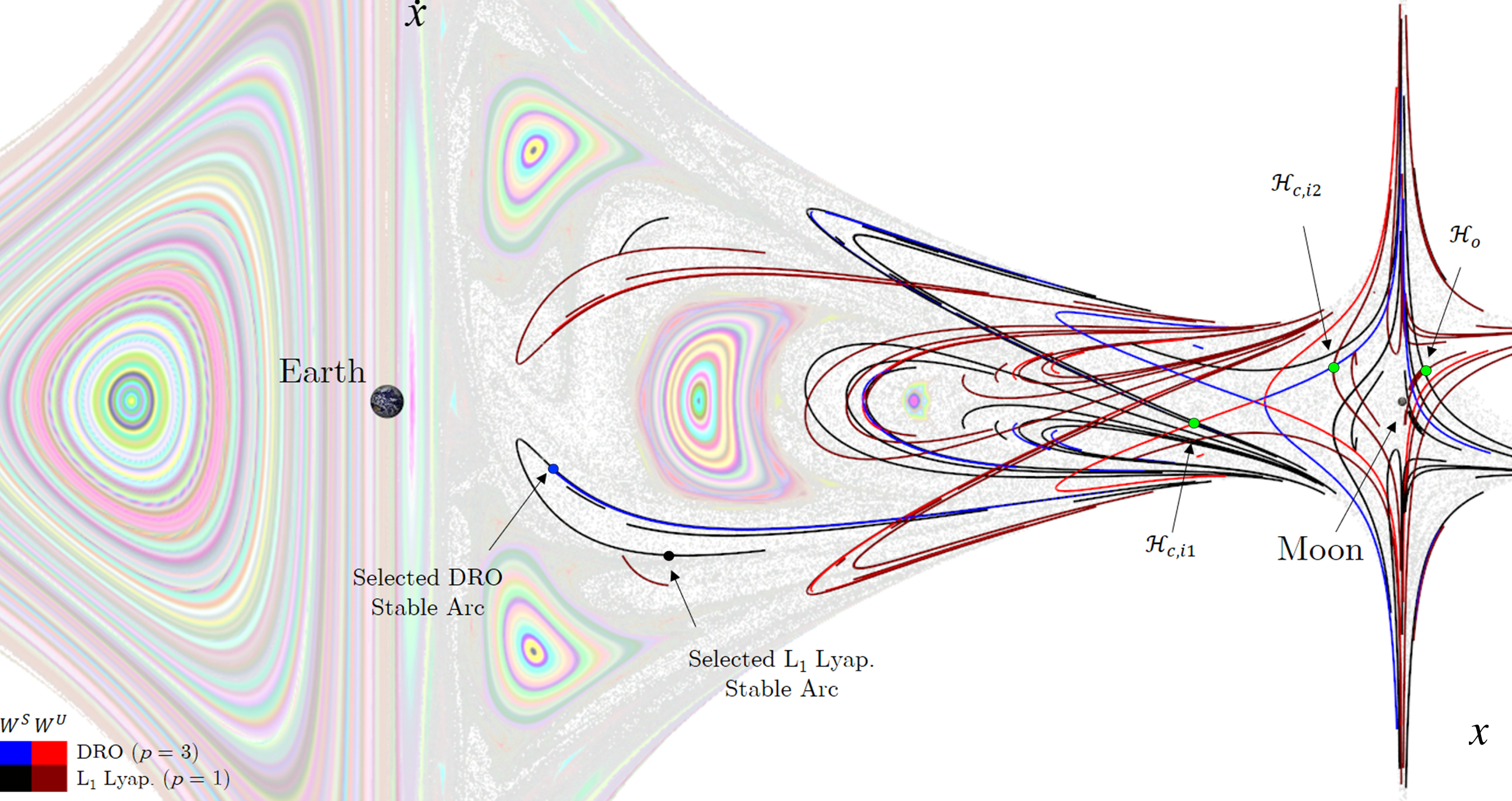}}
   \subfigure[Selected arcs (rotating frame)]{\label{sf:interLyapDROArcs}\includegraphics[width=0.49\linewidth]{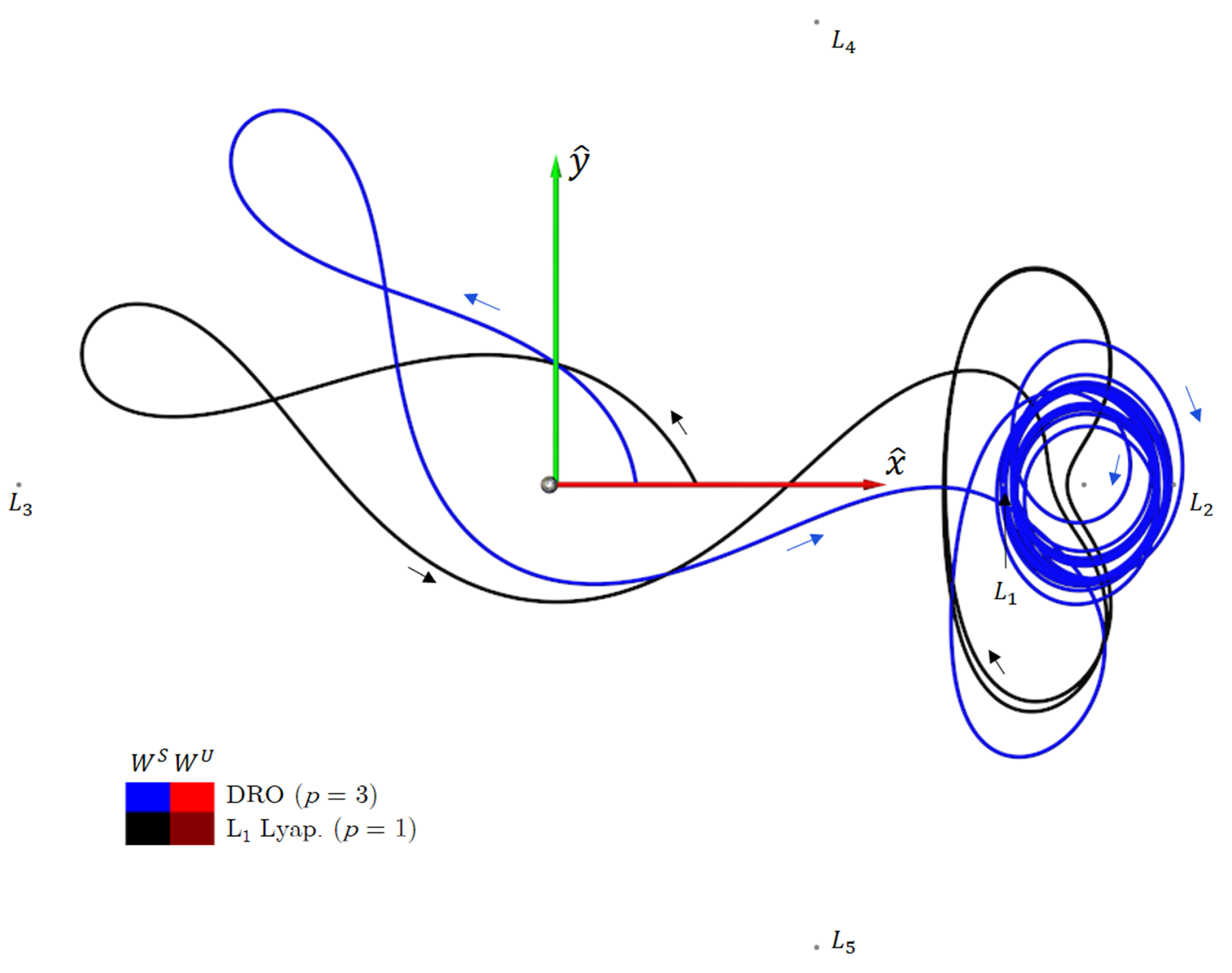}}
   \subfigure[Selected arcs (inertial frame)]{\label{sf:interLyapDROArcsInert}\includegraphics[width=0.49\linewidth]{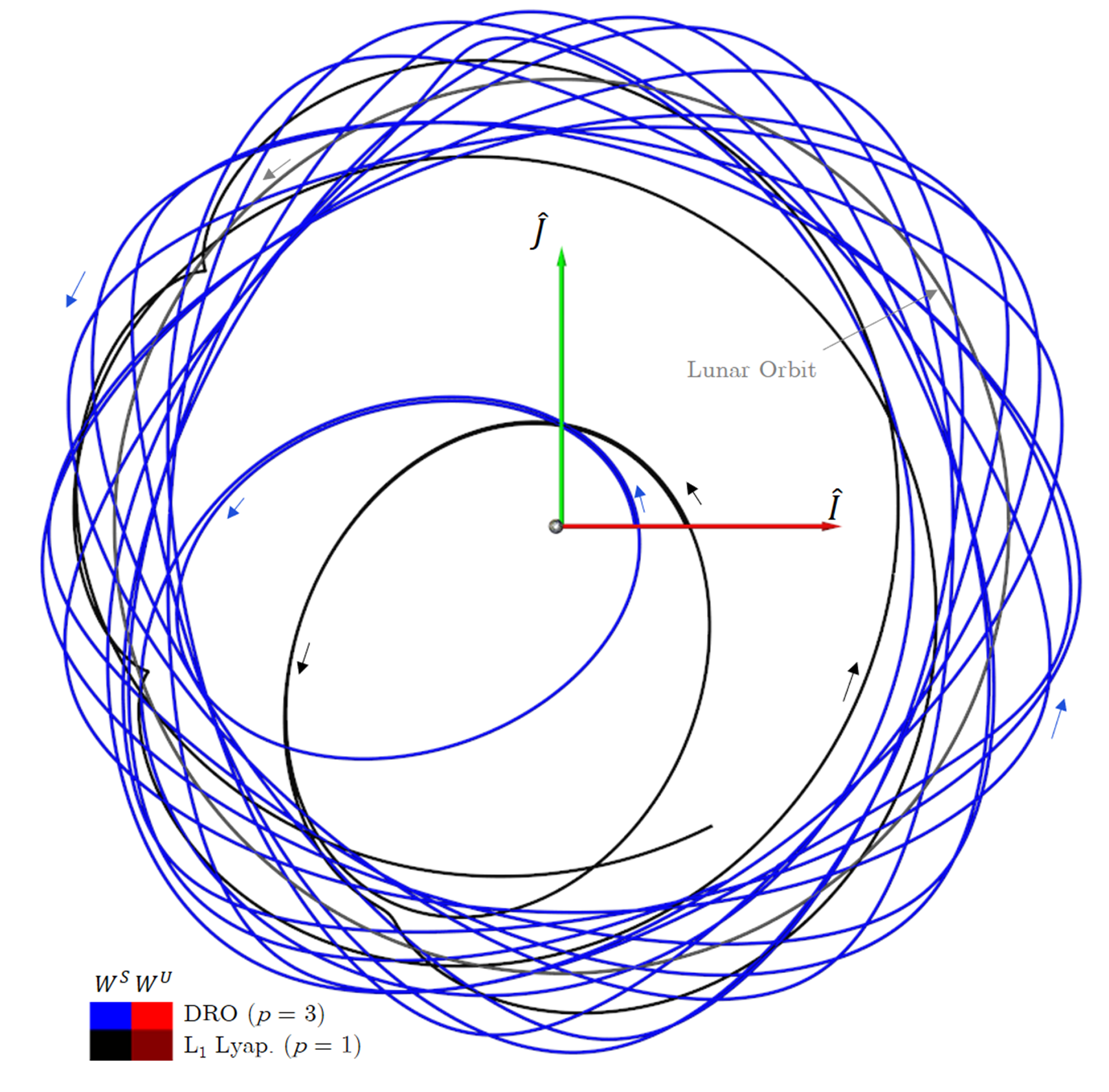}}
 \caption{Invariant manifolds for the $L_1$ Lyapunov and \dro saddle-type orbits extracted in the Earth-Moon system ($C=2.96$).  Selected stable manifold arcs are displayed in the rotating (b) and inertial (c) frames.}
 \label{fig:emInterLyapDRO}
\end{figure}

To illustrate this process, we first consider two simple transfers, namely from Earth proximity to the $L_1$ Lyapunov orbit (which surrounds $L_1$, see \autoref{fig:cr3bp_explanation}) on one hand, and to the distant retrograde orbit (DRO) of period $p=3$, on the other hand. Note that both targeted orbits are unstable, \ie{} their corresponding fixed points are saddles. We start by visualizing their invariant manifolds (see \autoref{sf:interLyapDROMap}). Note that the fragmentation of the manifold is due to transversality violations. The stable and unstable manifold pairs are colored with black and crimson for the $L_1$ Lyapunov and with blue and red for the \droNS, respectively. The basic idea in this case consists in letting the spacecraft follow the \emph{stable} manifold of the targeted orbit since the dynamics along that manifold is known to converge toward the orbit, by definition (see \autoref{fig:topology}).

Using this filtered visualization of the topology, the user can easily select entry points on a $L_1$ Lyapunov stable manifold (black) and on a \dro stable manifold arc (blue) at the indicated locations (\autoref{sf:interLyapDROMap}). Using the arc information recorded during the construction of these manifolds (see \autoref{sec:manifold}), our program automatically determines the path associated with each selection. The corresponding trajectories (shown in the rotating frame in \autoref{sf:interLyapDROArcs} and in the inertial frame\footnote{An inertial frame is a reference frame, \eg{} defined with respect to fixed distant stars, in which the rotating frame rotates.} in \autoref{sf:interLyapDROArcsInert}) initially perform an elliptical orbit around the Earth before progressively shifting towards an asymptotic approach of the targeted orbits. An important practical question from a design standpoint is the duration (or \emph{time of flight}) of each transfer. The black arc enters the $L_1$ Lyapunov orbit after 38.10 days whereas the blue arc arrives in the \dro after 88.14 days. In both cases, the convergence was determined within a small distance tolerance~\cite{Schlei:2017:Interactive}. Clearly, the arc selection capability enhances quick design construction by revealing many ballistic (\ie{} maneuver-free) capture trajectories.

\label{ssec:hhCon}

\begin{figure}[ht]
 \centering
   \subfigure[Manifolds for the \dro and Orbit $O^*$]{\label{sf:hhDROto207Map}\includegraphics[width=\linewidth]{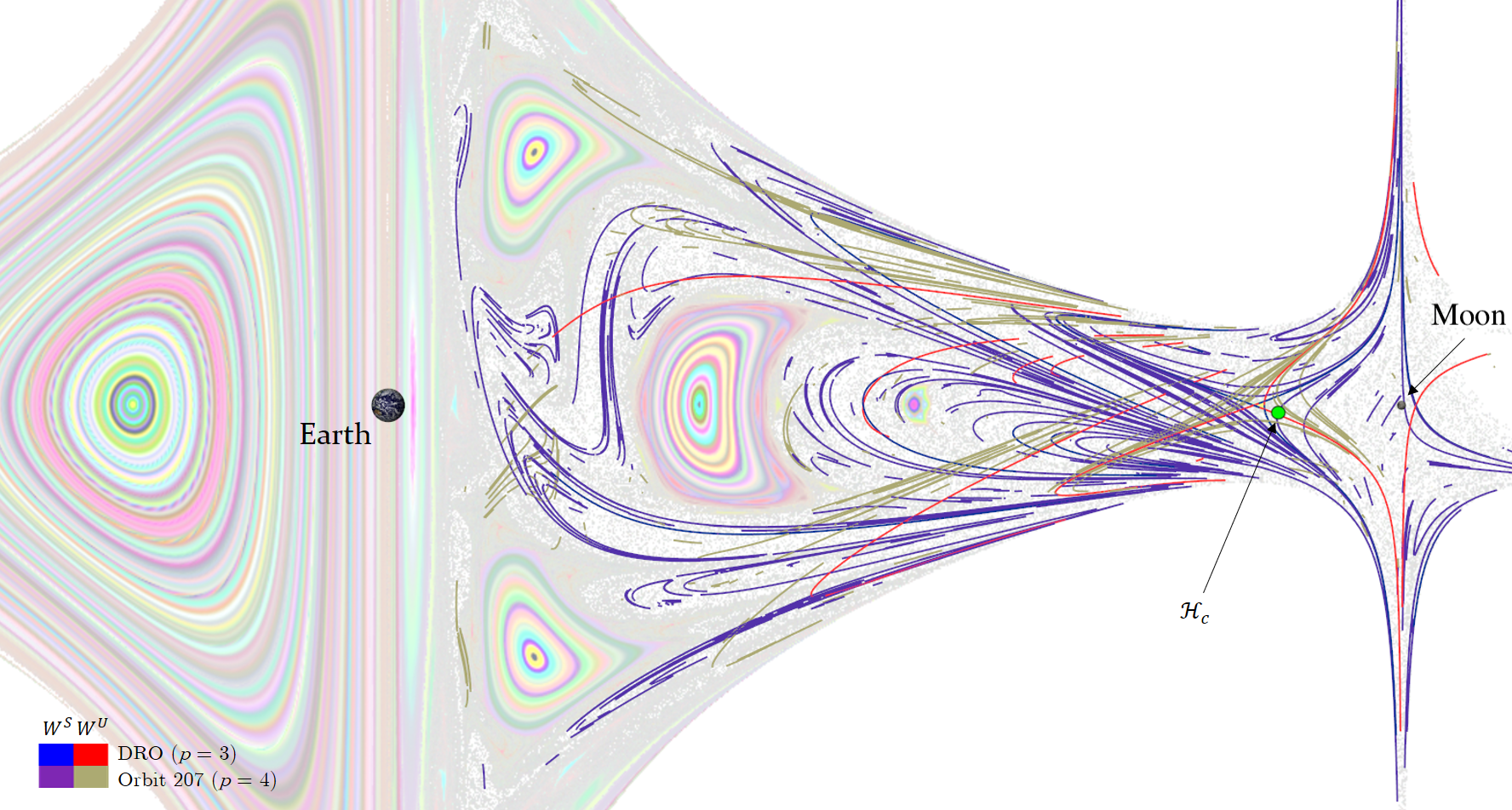}}
   \subfigure[$\mathit{H}_{c}:$ \dro to Orbit $O^*$ ($\Delta t = 197.22$ days)]{\label{sf:hcDROto207}\includegraphics[width=\linewidth]{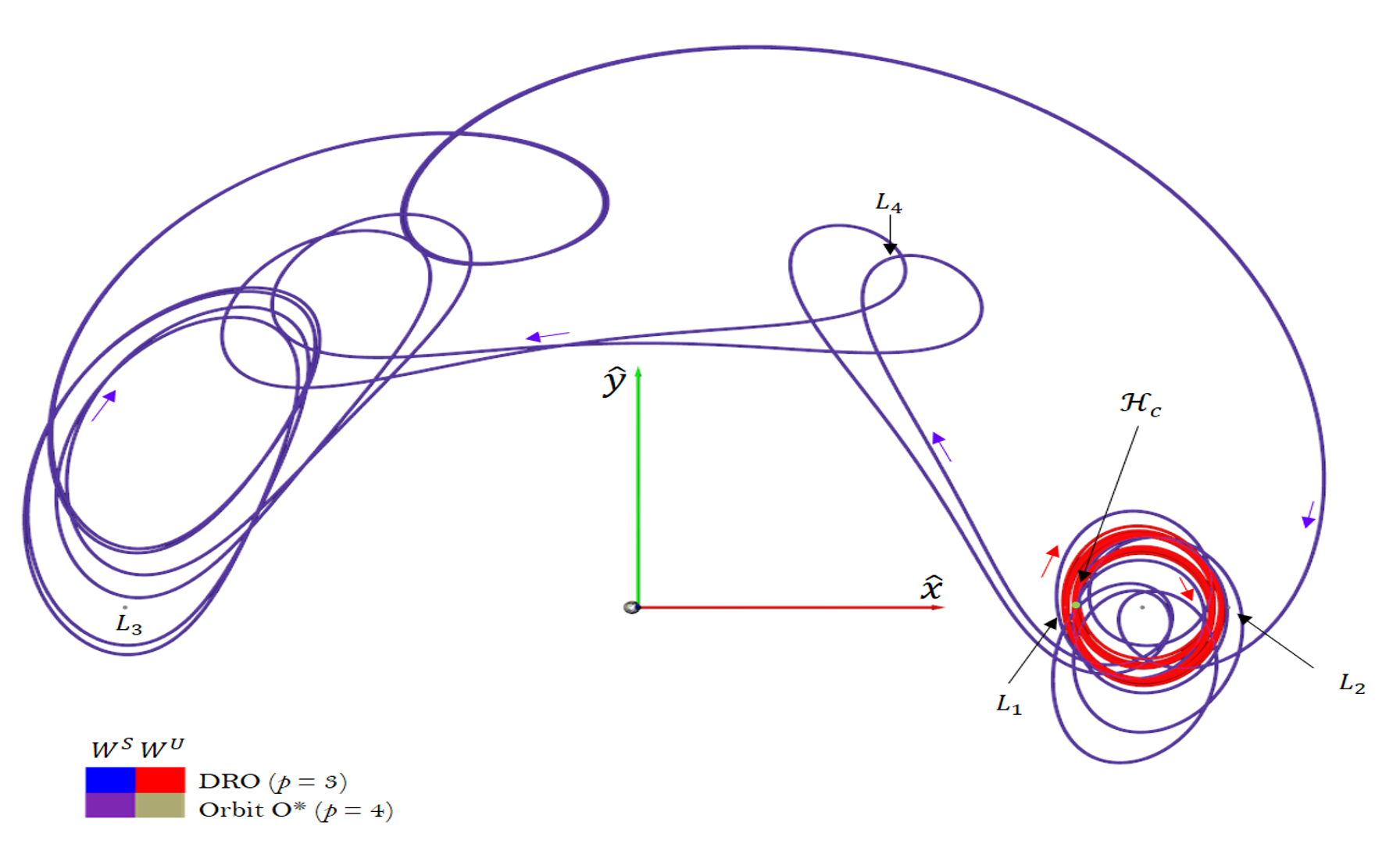}}

 \caption{Invariant manifolds for the \dro and Orbit $O^*$ extracted by our method and a selected heteroclinic connection in the Earth-Moon system ($C=2.96$).}
 \label{fig:hcDROto207}
\end{figure}

As second example, we now consider an interesting transfer starting at the \dro orbit mentioned above and departing to the rather convoluted orbit that visits $L_3$ and $L_4$ vicinities while also closely approaching the Moon several times.
We refer to this orbit as \emph{Orbit $O^*$} in the following. The visualization of the invariant manifold curves of that orbit (displayed in \autoref{fig:hcDROto207} with indigo and tan colors) reveals to the user that a lot of locations naturally flow into this orbit with the large dispersion of stable manifolds.  A transfer from the \dro to this orbit is represented by any red-indigo intersection. Among them, a selected option documents a transfer possibility as shown in \autoref{sf:hcDROto207}.
The wealth of geometric detail conveyed by our manifold visualization combined with the ability of our solution to explicitly construct heteroclinic connections make the selection of a transfer candidate quite straightforward. In addition, in the present case, the nearly space-filling nature of the manifold gives the user enormous spatial flexibility in the selection of the connection, which is very desirable from a path planning standpoint.

\subsection{Access to Enceladus}\label{sec:EncPath}

A popular topic in the astrodynamics community is the design of a low-cost pathway to an orbit around a moon of Saturn called Enceladus. A tantalizing scenario in that regard would be a fully ballistic capture, namely a solution in which gravity alone would deposit the spacecraft into a close orbit around Enceladus. Fortunately, our topology visualization makes it easy to construct such a path, even without detailed knowledge of the Saturn-Enceladus (\textit{SEnc}) system beforehand.

Practically, we select an energy level $C=2.999995$, which is known to contain a $p=3$ DRO, similar to the one considered previously in the Earth-Moon system, that is close to Enceladus (roughly 600 km) and permits safe transfer.

\begin{figure}[ht]
 \centering
    \subfigure[\Poincare section in Enceladus area]{\label{sf:sencMans2}\includegraphics[width=0.85\columnwidth]{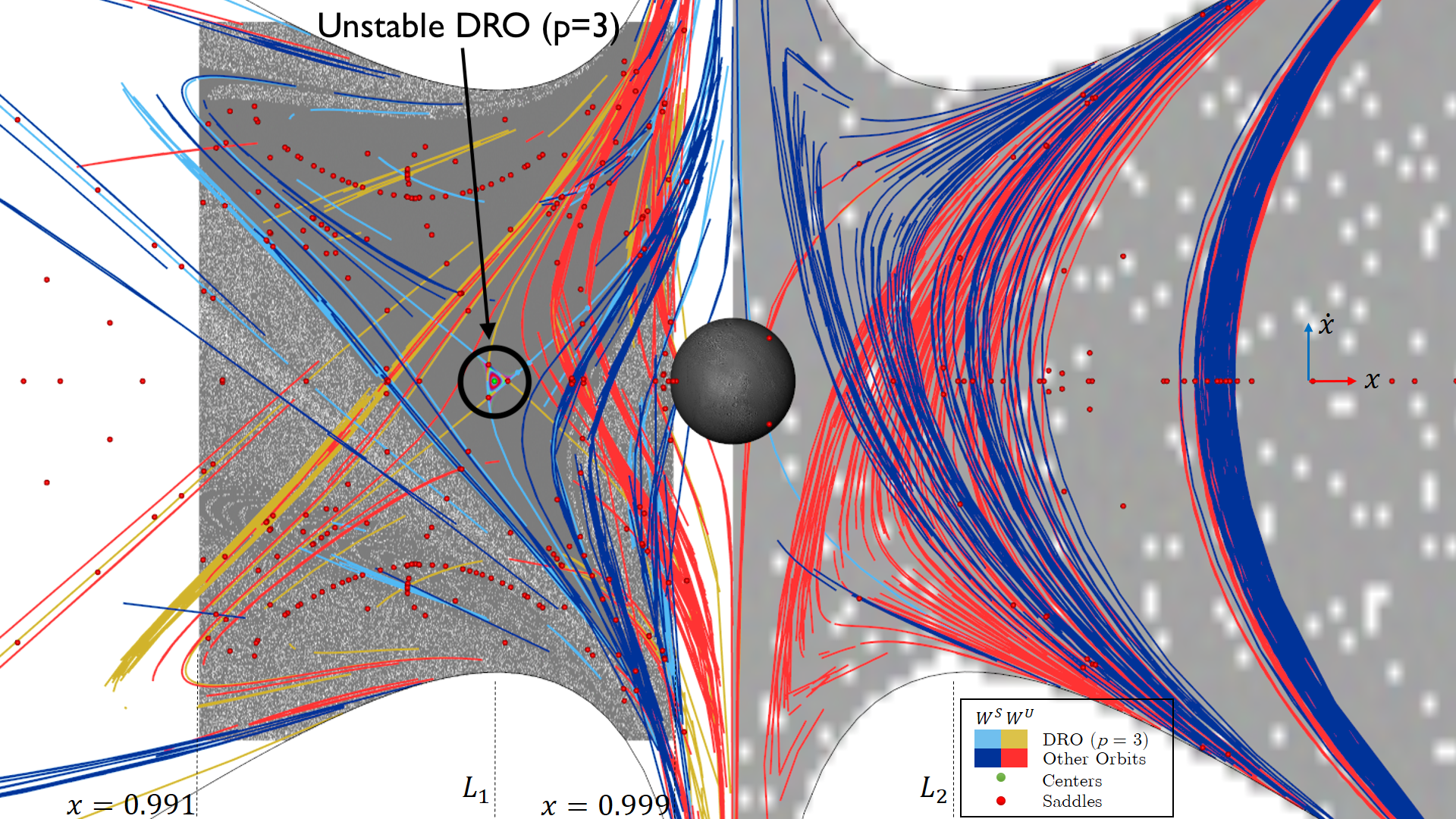}}
 \subfigure[\Poincare section near $L_3$]{\label{sf:sencBCZ}\includegraphics[width=0.85\columnwidth]{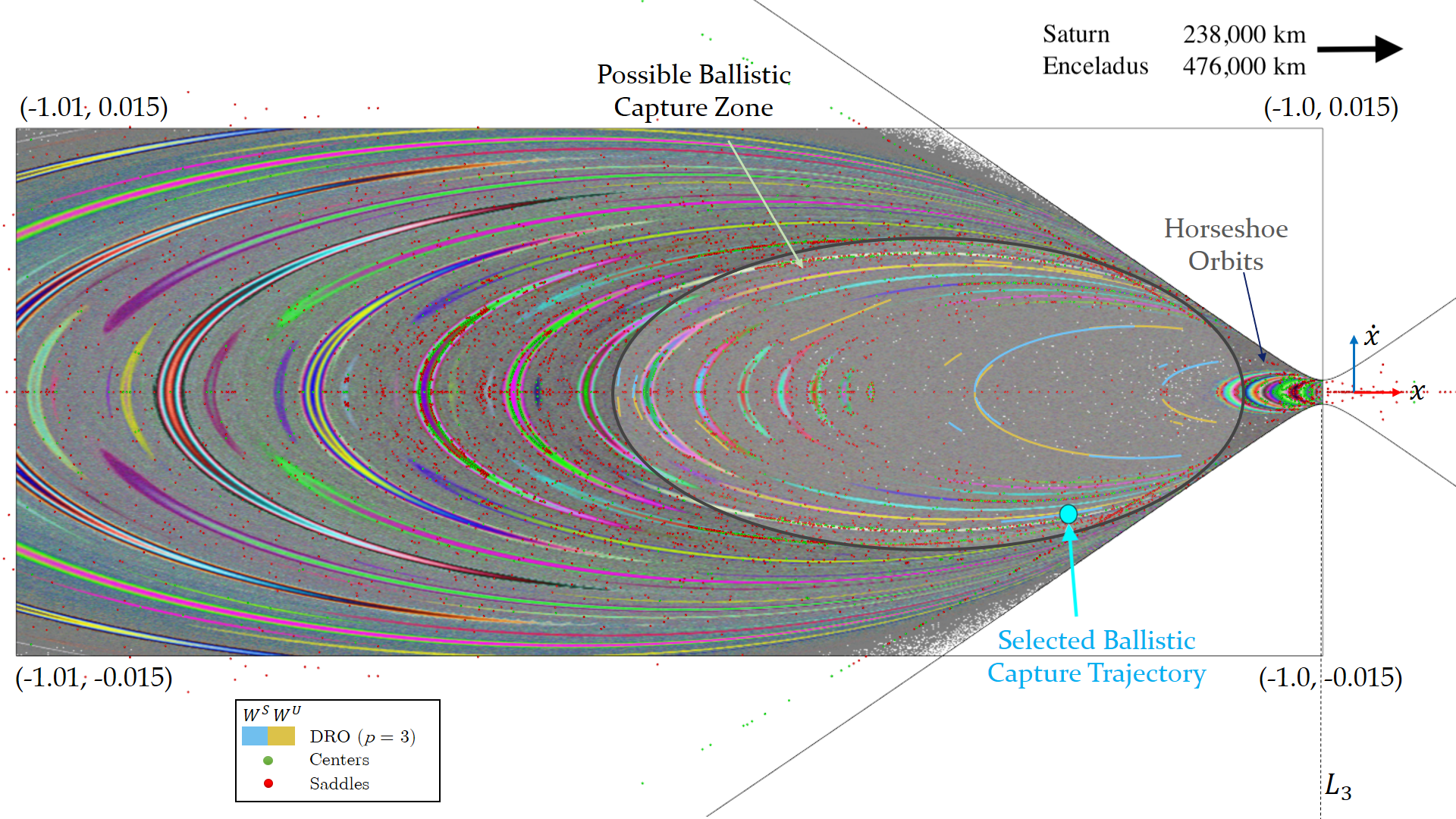}}
   \subfigure[Selected capture arc (\dro $W^S$)]{\label{sf:sencDROArc}\includegraphics[width=0.85\columnwidth]{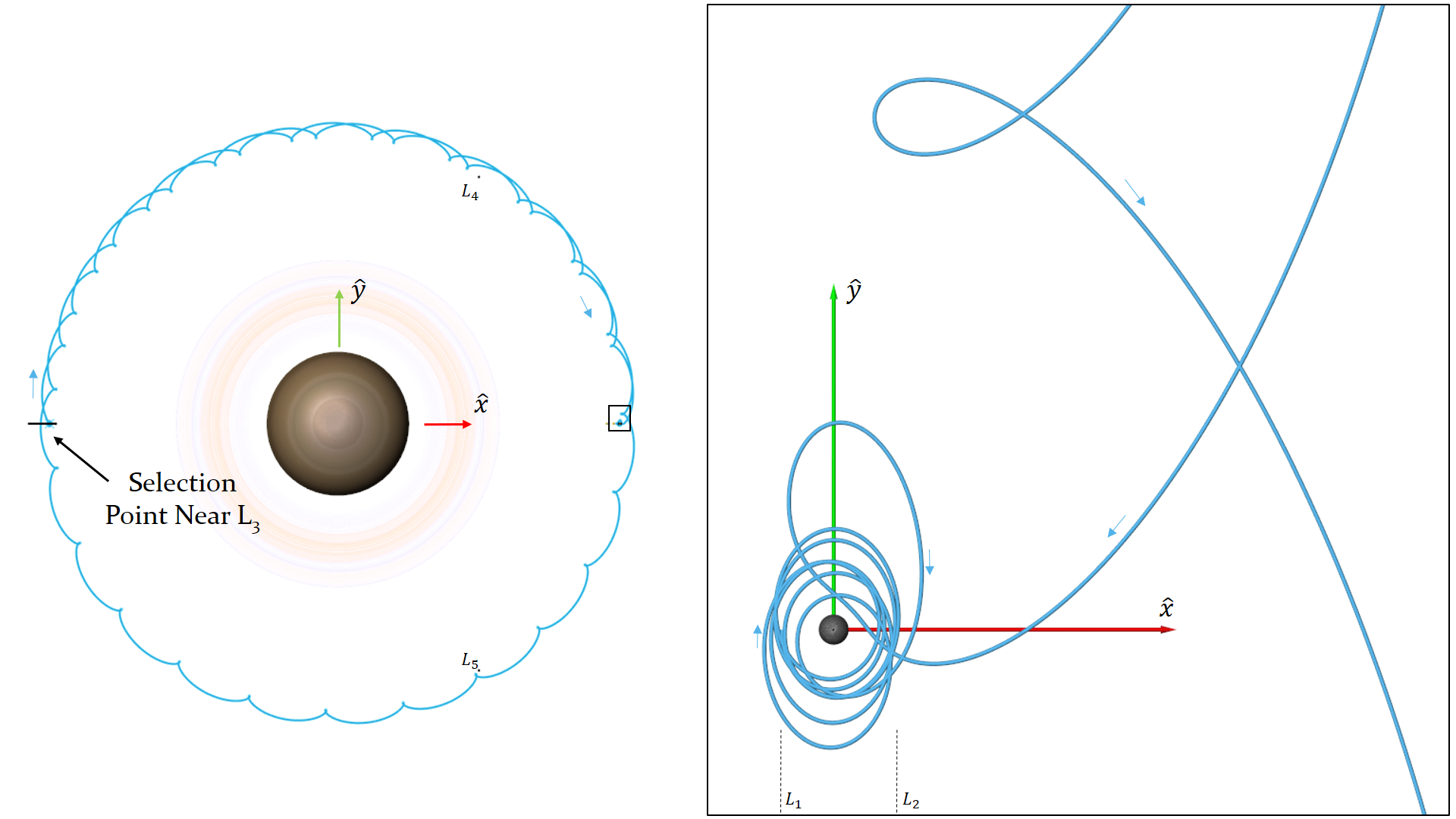}}
 \caption{A plausible region for ballistic capture about Enceladus indicated on a \Poincare section with a selected example \dro manifold arc in the \textit{SEnc} system at $C=2.999995$. }
 \label{fig:sencBCZ}
\end{figure}

Our method is employed to extract fixed points and invariant manifolds throughout the \textit{SEnc} system. The corresponding visualization is shown in \autoref{sf:sencMans2}. As in our first example (\autoref{sec:design}), we are here again seeking a transfer to a stable manifold of the DRO to achieve guaranteed convergence. This manifold, shown in light blue, exhibits many transfer options in the form of intersections with the dark blue unstable manifolds of the many saddle-type periodic orbits that populate that region. It can further be seen that these manifold segments populate not only the immediate vicinity of Enceladus but also a region situated on the opposite side of Saturn, shown in \autoref{sf:sencBCZ} over an orbital convolution texture~\cite{Tricoche:2011:Visualization} for context.

A possible solution consists in selecting the position shown in \autoref{sf:sencBCZ} where our topology visualization indicates the presence of an intersection between the unstable manifold of another periodic orbit and a segment of the DRO stable manifold on the opposite side of Saturn~\cite{Schlei:2017:Interactive}. Note that the dark blue unstable manifolds shown in \autoref{sf:sencMans2} densely cover this region as well but are omitted in the second image for clarity. The trajectory associated with this selection is shown in \autoref{sf:sencDROArc}. It can be seen that it performs one and a half revolutions around Saturn before a close passage of Enceladus that precedes the expected asymptotic approach of the DRO. Note that this particular solution is only one out many alternative options that our topology visualization automatically reveals, here again affording the designer precious flexibility in their selection of a suitable path.

\section{Discussion and Conclusion}
We have presented a \Poincare map topology visualization method that allows space trajectory designers to interactively discover novel natural transfer options in the circular restricted three-body problem as they piece together the path of a spacecraft in a planned mission. To make this visualization possible, we have devised a technical solution for the extraction of the \Poincare map topology in the CR3BP that allows us to identify thousands of fixed points and associated manifolds, the overwhelming majority of which were previously unknown. We have shown how a detailed visualization of the topological skeleton reveals crucial information to domain experts as they seek low-propellant trajectories. In particular, we have presented representative design scenarios in which the knowledge of the heteroclinic connections that exist between the invariant manifolds of unstable periodic orbits provided the domain expert with the essential insight needed to construct suitable transfer solutions. Our method, by displaying the connectivity of orbital structures, offers spacecraft trajectory designers a broad range of interactive options without external computation. Our collaborators' experimentation with this approach suggests that a designer could employ automated topological skeletons as a visual input catalog from which pathways can be interactively selected, visualized, and assessed.

Our approach and its current implementation have some limitations that we wish to address. First, the computational cost of our topology extraction method is significant. While our fixed point extraction technique is essentially an embarassingly parallel task that can be completed within a few minutes, the individual manifold construction procedure is an expensive operation that can take many hours to finish if the numerical parameters that control its behavior are not chosen properly. We provide in appendix the parameter values that we have found to strike a good balance between accuracy and computation time. A more principled way to choose these parameters, or alternatively, a way to predict their impact on the compute time would greatly benefit this work. A second important issue concerns the accuracy of our evaluation of the \Poincare map. By definition, each iteration of this map requires the integration of a strongly nonlinear differential system, which is an inherent source of error. It would be important to quantify the uncertainty that it introduces in our characterization of the topology and to convey it visually to the designer. Another outstanding problem is the visual complexity of the topology and the difficulty to effectively manage it to allow the spacecraft trajectory designer to quickly access the relevant information. We have so far dealt with this issue by applying, after the fact, stability thresholds to fixed points and manifolds but more sophisticated and discriminative filtering criteria would bring a major improvement. In fact, it would be most desirable to be able to apply this kind of filtering in a pre-processing step in order to limit the scope of the computation to the most relevant features of the topology for the considered problem.

\acknowledgments{
This research was supported in part by NSF CAREER Award \#1150000. This work also benefited from a gift by Intel. The authors are grateful to Rune and Barbara Eliasen for their support funding the Eliasen Visualization Laboratory at Purdue University.  Also, the authors wish to acknowledge Visualization Sciences Group (an F.E.I. company and the developers of \Avizo) for implementation assistance with the visualization tools employed in this work.
}

\bibliographystyle{abbrv}
\bibliography{xmt}

\end{document}